%% file: main.tex
\documentclass{article}

\usepackage{microtype}
\usepackage{graphicx}
\usepackage{cancel}
\usepackage{subfigure}
\usepackage{booktabs} %

\usepackage{hyperref}
\usepackage{amsmath,amsthm,amsfonts,amssymb}
\usepackage{notation}

\usepackage[accepted]{icml2024}

\usepackage{amsmath}
\usepackage{amssymb}
\usepackage{mathtools}
\usepackage{amsthm}
\usepackage{bbm}
\usepackage{float}
\usepackage{dsfont}
\usepackage[capitalize]{cleveref}

\theoremstyle{plain}
\newtheorem{theorem}{Theorem}[section]
\newtheorem{proposition}[theorem]{Proposition}
\newtheorem{lemma}[theorem]{Lemma}
\newtheorem{corollary}[theorem]{Corollary}
\theoremstyle{definition}
\newtheorem{definition}[theorem]{Definition}

\theoremstyle{remark}

\icmltitlerunning{\papertitle}

\renewcommand{\Cref}[1]{\cref{#1}}

\begin{document}

\twocolumn[
\icmltitle{\papertitle}

\begin{icmlauthorlist}
\icmlauthor{Haiqing Zhu}{anu}
\icmlauthor{Alexander Soen}{anu,riken}
\icmlauthor{Yun Kuen Cheung}{anu}
\icmlauthor{Lexing Xie}{anu}
\end{icmlauthorlist}

\icmlaffiliation{anu}{School of Computing, The Australian National University, Canberra, Australia}
\icmlaffiliation{riken}{RIKEN Center for Advanced Intelligence Project, Tokyo, Japan}

\icmlcorrespondingauthor{~}{\{haiqing.zhu,  alexander.soen, yunkuen.cheung, lexing.xie\}@anu.edu.au}
\icmlkeywords{Online Learning, Regret, Betting Market, Prediction Market}

\vskip 0.3in
]

\printAffiliationsAndNotice{} %

\input{sections/Abstract}

\input{sections/Introduction}

\input{sections/Model}

\input{sections/Equilibria}

\input{sections/SA}

\input{sections/FTL}

\input{sections/Empirical}

\input{sections/Conclusion}

\ifnum\thepage=9
    \pagebreak
\fi

\input{sections/ack}

\input{sections/ImpactStatement}

\bibliography{refs}
\bibliographystyle{icml2024}

\newpage

\input{appendix}

\end{document}

%% file: sections/Abstract.tex
\begin{abstract}
We examine two types of binary betting markets, whose primary goal is for profit (such as sports gambling) or to gain information (such as prediction markets). We articulate the interplay between belief and price-setting to analyse both types of markets, and show that the goals of maximising bookmaker profit and eliciting information are fundamentally incompatible. 
A key insight is that profit hinges on the deviation between (the distribution of) bettor and true beliefs, and that heavier
tails in bettor belief distribution imply higher profit.
Our algorithmic contribution is to introduce online learning methods for price-setting. Traditionally bookmakers update their prices rather infrequently, we present two algorithms that guide price updates upon seeing each bet, assuming very little of bettor belief distributions. The online pricing algorithm achieves stochastic regret of $\bigoh(\sqrt{T})$ against the worst local maximum, or \( \bigoh(\sqrt{T \log T}) \) with high probability against the global maximum under fair odds.
More broadly, %
the inherent trade-off between profit and information-seeking in binary betting may inspire new understandings of large-scale multi-agent behaviour.
\end{abstract}

%% file: sections/Introduction.tex
\negspacetxt
\negspace
\negspace
\section{Introduction}
\negspace
\negspace

Betting on future events is interesting for at least two reasons: to elicit and aggregate belief as done in online prediction markets, or to profit from correct forecasts as done in traditional betting market. 
There is a vast (albeit separate) literature about prediction and betting markets. A closer examination prompts us to ask three questions. Can one describe prediction market and traditional betting market using a shared mathematical model? Given the many empirical studies on bettor and bookmaker strategies 
in  betting markets, 
what is a theoretical model for setting the book? Also, can prediction-market-style online algorithms help profit-oriented bookmakers? %

\paragraph{Related work.}
Prediction markets~\cite{wolfers2006interpreting} have had great algorithmic success as information aggregators. 
Early development focused on how market maker used \emph{proper scoring rule} to reward forecasters and encourage 
honest predictions~\cite{Brier1950,Good1952,McCarthy1956,Savage1971,Gneiting2007}.
A landmark result is the automated market maker \cite{hanson2007logarithmic} using the \emph{logarithmic market scoring rule} (LMSR).
The inherent connection of LMSR to online optimisation 
led to various generalisations of LMSR in algorithms and market structures \cite{ChenPennock2007,ChenGP2008,AgrawalWY2008,GuoP2009,ChenV2010,Agrawal2011OR,abernethy2013efficient,Frongillo:2012,Othman2013Liquidity}, and convergence analyses of the relevant dynamic processes~\cite{FrongilloR2015,Yu2022EC}.

For betting markets,
Kelly betting \cite{kelly1956new,rotando1992kelly,busseti2016risk} 
is a prominent model and applicable to
different domains \cite{thorp2008kelly}.
\citet{levitt2004gambling}'s empirical work on bookmaker strategy identified that better prediction ability %
and their systematic exploitation of bettor biases as two main drivers of bookmakers' profit.
 More recently, \citet{yu2022betting} characterised the uniqueness %
of the equilibrium of the betting market model under the special case that the odds are fair and no profit is earned.
Compared to the above, our work provides theoretical explanations of Levitt's findings based on a more general online betting model.
Furthermore, we show that bookmakers can earn higher profit not only by exploiting bettor biases, 
but also by exploiting polarised bettor belief distributions that have heavy tails.
Remarkably, we show that such exploitation is not only feasible, but can be done in an online setting with little knowledge of bettor belief distribution beforehand.

Our profit-maximisation problem is distinct from other online learning problems.
For example, multi-armed bandits \cite{thompson1933likelihood,slivkins2019introduction,lattimore2020bandit} optimise profit for the (repeating) bettor when the rate of return is unknown, whereas our problem maximises the bookmaker profit when bettor belief distribution is unknown.
The unknown belief distribution constitute the uncertain component in both the objective function and its gradient -- but noisy estimates of both are used in our algorithms. Focusing on gradient information, one could think of this problem as between Online Convex Optimisation~\cite{hazan2016introduction} with known gradient and Bandit Convex Optimisation problems~\cite{hazan2014bandit,lattimore2024bandit} with no gradient available.
Additionally, the optimisation objective here is also uncertain and generally non-convex.

\paragraph{Our contributions.} 
We consider markets betting on binary outcomes. \cref{table:teaser} summarises several key concerns. Two choices determine the market setting: whether the bookmaker aims to update their own belief -- {\it No (N)} for traditional betting, and {\it Yes (Y)} for prediction markets; and whether prices for the two outcomes are {\it fair}. Here fair odds/prices {\it Y} means the implied event probabilities sum to one, and unfair odds {\it N} means the bookmaker ignore this constraint to ensure profit.
Traditional betting markets can seek to maximise profit with either fair or unfair odds, whereas prediction market seek to aggregate belief and invariably uses fair odds tied to the estimates of the event probability. The setting of belief update under unfair odds has contradictory effects -- namely, discouraging some bettor from participating
but trying to elicit their belief -- and therefore is out of scope from this (and others') work. This work assumes the common Kelly model for bettor behaviour, which allows us to focus on the properties and strategies of the bookmaker. 

\input{figure/teaser}

Our first set of results describe the conflict between profit and prediction (in \cref{sec:model,sec:equilibria}). We show that bookmakers can expect positive profit in both fair and unfair odds settings. In particular, 
the bookmaker setting a price to exploit the difference between the bettors' average belief and their own will lead to profit.
On the other hand, when the bookmaker set their own belief to match the bettors' average belief, the setting turns into a prediction market with maximum profit being zero.
Moreover, we show that second order stochastic dominance relations across bettor belief distributions implies the same ordering in profit -- the bookmaker can obtain more profit from heavier tails of the distributions.
To the best of our knowledge, this is the first formal result on bookmaker profit under unfair odds. 

The next contribution is to introduce online learning to book-making, resulting in two algorithms that optimise profit. 
\cref{alg:SA} (\cref{sec:sa}) is a %
stochastic approximation algorithm~\citep{mertikopoulos2020almost} to find optimal prices with unfair odds. It does so with local gradient updates, and achieves a regret of \( \bigoh(\sqrt{T}) \) against %
a local maximum.
\cref{alg: FTL} (\cref{sec:ftl}) is a \emph{Follow the Leader}~\citep{hazan2016introduction}-type algorithm 
in the fair odds setting. It computes a running estimate of average bettor belief and uses the first-order analytical solution in the fair odds setting to converge to globally optimal prices.
We observe both algorithms outperform the respective risk-balancing and logarithmic market scoring baselines in empirical simulations, and are robust to different initialisations (\cref{sec:Empirical}).

%% file: figure/teaser.tex
\begin{table}[t!]
\bignegspace
 \caption{An overview of market settings and key quantities in online learning. Our contributions are highlighted in {\color{olive} green}. 
 $*$ with high probability.
 $\#$ the algorithm is due to~\cite{Frongillo:2012}. 
 $\dagger$ captures local maxima and has stronger assumptions.}
 \resizebox{\columnwidth}{!}{
     \centering
    \begin{tabular}{cccccc}
    \toprule
    \multicolumn{3}{l}{Market Setting} & \multicolumn{3}{l}{Market Properties} \\
    \cmidrule(lr){1-3}
    \cmidrule(lr){4-6}
    Goal & \shortstack{Update \\ Belief?} & \shortstack{Fair \\ odds?} & \shortstack{Max \\ Profit } & \shortstack{Online \\ Learning} & Regret \\
    \midrule
    \multirow{2}{*}{Profit} & \multirow{2}{*}{N} & N & {\color{olive}$>0$} &{\color{olive}SA Sec \ref{sec:sa}} & {\color{olive}\( \bigoh(\sqrt{T}) \)}{\({}^\dagger\)} \\
     &  & Y & {\color{olive}$>0$} & {\color{olive}FTL Sec \ref{sec:ftl}} & {\color{olive}\( \bigoh(\sqrt{T \log T}) \)}\({}^*\) \\
     Prediction & Y & Y & $=0$ & Mirror Descent$^\#$ & \( \bigoh(\sqrt{T \log T}) \){\({}^*\)} \\
    \bottomrule
    \end{tabular}%
    \label{table:teaser}
    }%
\bignegspace
\bignegspace
\bignegspace
\negspace
\end{table}

%% file: sections/Model.tex
\section{Betting Markets}
\label{sec:model}
Consider a betting market on binary outcomes, called \eventR and \eventL\footnote{We choose to focus on pre-determined binary outcomes in this work, e.g. {\it home team winning} in a sporting event. Such setting is commonly used in prediction markets~\cite{beygelzimer12Learning,Frongillo:2012,wolfers2006interpreting}. We assume bookmakers are setting the price for \eventR and \eventL but not defining the event, such as the point spread at which home team winning~\cite{levitt2004gambling}. 
We leave non-binary betting markets as future work.
}. 
A {\it bookmaker} sets a price $\priceR$ for \eventR and $\priceL$ for \eventL, with $\priceR, \priceL \in (0, 1]$ , respectively\footnote{Here $\priceR$, $\priceL$ represent {\it price for unit return}. Another commonly used, and equivalent, representation is to express them in {\it odds}, which are $1/\priceR$ and $1/\priceL$, respectively.}. 
A {\it bettor} bets an amount $v$
on either outcome. If they {\it lose}, $v$ goes to the bookmaker; if they {\it win}, ${v}/{a}$ or ${v}/{b}$ is paid out to the bettor.

Denote the bookmaker's estimate of the
underlying probability of event \eventR as $g$.%
If the bookmaker and the bettors all took $g$ as their beliefs that \eventR occurs,
then setting prices $\priceR =g$ and $\priceL =1-g$ would {\it clear} the market, with no one expecting to make a profit -- but this setting is neither realistic nor interesting.
The rest of this section considers the bettor-bookmaker dynamic as a game, and addresses the following questions. If the individual belief of the $t$-th bettor of \eventR is $p_t \in [0, 1]$, then how much would they bet given the prices $(\priceR, \priceL)$? Given such utility-maximizing bettors, what is the expected profit for the bookmaker?

A table of key notation is presented in \cref{app:notation}.
Proofs and additional expositions are deferred to the appendix.

\negspace
\bignegspace
\subsection{Kelly bettors and their optimal strategies}
\label{ssec:model_bettor}
\negspace
\negspace
We consider \emph{Kelly bettors}~\citep{kelly1956new}, 
indexed sequentially by $t \in \{1,\ldots,T\}$.
Each bettor possesses an initial wealth \( \wt \) and has an underlying belief \( \pt \) that event \eventR will occur, implying that belief for event \eventL is \( \qt \defeq 1 - \pt \).

A Kelly bettor bets an amount $\bet_t$
that maximises a utility representing the expected logarithm of their wealth, %
\begin{equation}\label{eq:bettor-utility-function}
 \bet_t \defeq \argmax_{v \colon v\geq 0} \left\{ \varphi_t^{\priceR}(v) \vee \varphi_t^{\priceL}(v) \right\}.
\end{equation}
Here $\varphi_t^{\priceR}(v)$ and $\varphi_t^{\priceL}(v)$ are the expected log wealth after bettor $t$ bets on event \eventR or \eventL, respectively. The inner $\max$ (via symbol $\vee$) indicate which side to bet on, while the outer max is over the amount of the bet.
\bignegspace
\begin{align*}
    \varphi_{t}^{\priceR}(v) &= \pt \log\left(\wt + \frac{1-\priceR}{\priceR} v \right) + \qt\log(\wt - v ); \\
    \varphi_{t}^{\priceL}(v) &= \qt \log\left(\wt + \frac{1-\priceL}{\priceL} v \right) + \pt\log(\wt - v ).
\negspace
\negspace
\end{align*}
The first term in utility function $\varphi_t^\priceR$ is the probability of \eventR happening $\pt$ times the log of the resulting wealth, with the profit from the win being $\frac{1-\priceR}{\priceR}v$. The second term is the probability of \eventL happening $\qt$ times the log of initial wealth $\wt$ subtracting the loss $v$. Similarly for utility $\varphi_t^\priceL$ \wrt event \eventL.
\cref{eq:kelly-betting-rule} contains the analytic solution for Problem \eqref{eq:bettor-utility-function} where the first case corresponds betting on \eventR and the second case corresponds to betting on \eventL.
See \cref{appendix_kelly} for a derivation and an illustration.
\negspace
\begin{equation}
\bet_t = \begin{cases}
    \frac{\pt - \priceR}{1 - \priceR}\cdot \wt, &\text{if}~\pt>\priceR;\\
    0, &\text{if}~1-\priceL \leq \pt \leq \priceR;\\
    \frac{\qt - \priceL}{1 - \priceL}\cdot \wt, &\text{if}~\qt>\priceL.
\end{cases}\label{eq:kelly-betting-rule}
\end{equation}
The Kelly betting model is commonly applied in various fields, including sports betting and gambling \cite{thorp2008kelly}, portfolio management \cite{thorp1975portfolio}, and stock market \cite{rotando1992kelly}.
It is proven to be asymptotically optimal such that it maximises the long-term compounded returns~\cite{kelly1956new, algoet1988asymptotic,cover1999elements} . Despite its effectiveness, the Kelly betting approach faces criticism for being overly aggressive \cite{busseti2016risk} or, conversely, too conservative \cite{hsieh2016kelly}. To address these issues, various alternative betting strategies have been proposed, such as fractional  \cite{davis2013fractional} and distributional robust \cite{sun2018distributional} Kelly and modern portfolio theory \cite{markowitz1952portfolio}. Our model is applicable for fractional Kelly strategy by replacing the wealth distribution with a joint distribution of wealth and fractions. Detailed analysis of these alternative strategies is left for future research.%

\subsection{The bookmaker: maximising expected profit}
\label{ssec:model_bookmaker}
Key moderation signals for betting markets are the prices $(\priceR, \priceL)$, set by the bookmaker. The setting when $\priceR+\priceL=1$ is called {\it fair} in prices and odds~\cite{wiki-mathbookmaking}\footnote{$a+b<1$ allows arbitrage -- guaranteed profit for bettors placing bets on both sides simultaneously -- and hence not applicable.};
the setting $\priceR+\priceL>1$ is commonly used to guarantee bookmaker profits. 
The amount  $\priceR+\priceL-1$  is called {\it overround} %
or bookmaker {\it margin}. 
Given prices $(\priceR, \priceL)$ and a bookmaker's belief $g$ of event \eventR happening, one can calculate {\it their} expected profit $\ut(\priceR, \priceL)$
for a single bettor with unit wealth ($\wt=1$), where the expectation is taken over bettor belief $\pt$:
\begin{equation}
    \begin{aligned}
        \ut(\priceR, \priceL) \defeq \left(\frac{1-g}{1-\priceR} - \frac{g}{\priceR}\right)&\cdot \expect{(\pt -\priceR)_{+}} \\
        + \left(\frac{g}{1-\priceL} - \frac{1-g}{\priceL}\right)&\cdot \expect{(\qt -\priceL)_{+}}.  \label{eq:profit-cont}
    \end{aligned}
\end{equation}
Derivation of \cref{eq:profit-cont} is in \cref{app: derivation of profit}, 
which involves breaking down expected payin minus payout according to event \eventR or \eventL actually happening using \cref{eq:kelly-betting-rule}. 
For both terms in \cref{eq:profit-cont} to be non-negative, the bookmaker would set \( \priceR \in [g, 1] \) and \( \priceL \in [1-g, 1] \). \cref{eq:profit-cont} is inherently connected to the notion of \emph{conditional value-at-risk} (CVaR)~\cite{rockafellar2000optimization} in financial markets due to both terms being functions of conditional expectations \wrt the tails of belief distributions.
An exposition of which is presented in \cref{sec:app_cvar}.
Note that \cref{eq:profit-cont} does not use the true event probability of \eventR, but only the bookmaker's belief $g$, which leads to the following discussion.
\bignegspace
\bignegspace
\paragraph{The bookmaker's belief versus event probabilities.} $g$ is a key quantity that influences the expected profit. In real-world markets such as sports betting, bookmaker belief is known to be at least as close to the true probability as the most competent bettor~\cite{levitt2004gambling}. 
Nonetheless, $g$ is {\it not} the true probability of event \eventR.
\cref{eq:profit-cont} is still meaningful because it is an expectation from the bookmaker's perspective.
Maximising \cref{eq:profit-cont} is also theoretically and practically sound, because even when the bookmaker's belief is uncertain, i.e., in the form of an interval $(g_-, g_+)$ enclosing the ground-truth probability, we show that bookmaker profit under true event probability is still positive, and that algorithms proposed in \cref{sec:sa} can maximise its lower bound with trivial changes to incorporate $g_-, g_+$.
Details of this \emph{imprecise belief lemma} are in \cref{app:imprecise_belief}.

\paragraph{Profit under fair odds and prediction market.}
When $\priceR + \priceL = 1$, the expected profit function \cref{eq:profit-cont} simplifies to
\negspace
\negspace
\begin{equation}
\label{eq:fair_profit}
u_t(\priceR, 1-\priceR) = - \frac{(\priceR - g)(\priceR-\expect{\pt}) }{\priceR(1-\priceR)}.
\bignegspace
\end{equation}
Notice that the denominator is positive. %
Thus, the bookmaker makes a profit iff $\priceR$ is strictly between $g$ and $\expect{\pt}$.
In other words, the bookmaker's profit hinges on the aggregated bettor opinion deviating from their own belief $g$.

When the bookmaker's belief coincides with the aggregated bettor opinion, or \( g = \expect{\pt}\) under fair odds, 
the optimal market price for the bookmaker becomes exactly \( \priceR^{\star} = g \), leading to an expected profit of zero -- \ie, the price elucidates the mean belief of bettors. This includes the setting of prediction markets~\cite{beygelzimer12Learning,wolfers2006interpreting}.
This is surprising as the underlying dynamics of prediction markets optimise a utility function entirely different from the average profits. Indeed, the implicit utility function of the prediction market is \( \priceR \mapsto -\kl(\expect{\pt} \Mid \priceR) \)~\citep[Corollary 1]{Frongillo:2012}, which directly corresponds to information elucidation.

%% file: sections/Equilibria.tex
\section{Equilibria in Betting Markets}
\label{sec:equilibria}

Notice that \cref{eq:kelly-betting-rule} depicts the behaviours of bettors when they maximise their own utility; and \cref{eq:profit-cont} is the bookmaker's possible utility when bettors act in this optimal manner.
A natural question arises of whether there exists an equilibrium state where both the bettors and bookmaker's utility are maximised by a pair of prices $(\priceR^\star, \priceL^\star)$ and investments $\bet_t$ (for all $t \in \{ 1, \ldots, T \}$).

To consider this question, we first define the bookmaker's total utility over all bettors $\{1, \ldots, T\}$. 
This is simply defined as the  sum of \cref{eq:profit-cont} weighted by expected wealth/budget of each bettor:
\begin{equation}
    \label{eq:total_bookmaker_util}
    u_{1:T}(\priceR,\priceL) \defeq \sum_{t=1}^T \ut(\priceR,\priceL)\cdot \expect{\wt}.
\end{equation}
Here, we are using the assumption that the wealth $\wt$ and belief $\pt$ distributions for bettors are independent.
We further assume the \pdf of these distribution is differentiable almost everywhere. %
The function $u_{1:T}$ in \cref{eq:argmax-profit} is generally non-concave when $\priceR + \priceL > 1$ (via \cref{eq:profit-cont}). 

This utility function encodes a two-stage Stackelberg game \cite{von2010market} between the bookmaker and the bettors.
Indeed, with the knowledge that the bettors make their optimal response simultaneously, the bookmaker (as the Stackelberg leader) can make a pricing decision based on  \cref{eq:total_bookmaker_util}. When the bookmaker makes an optimal pricing, a Stackelberg equilibrium is achieved.
\begin{definition}
    A Stackelberg equilibrium 
    is achieved when the bettor with utility functions $\varphi^{\priceR^\star}$ and $\varphi^{\priceL^\star}$ bets optimally with wager from \cref{eq:bettor-utility-function}
    and the bookmaker sets the price to maximise the expected profit over a set of bettors assuming the bettors bet optimally
    \bignegspace
    \negspace
    \begin{equation}
        (\priceR^\star, \priceL^\star) = \argmax_{(\priceR,\priceL) \colon \priceR + \priceL \geq 1}~u_{1:T}(\priceR, \priceL). \label{eq:argmax-profit}
        \negspace
        \bignegspace
    \end{equation}\label{def: stackelburg-equilibrium}
\end{definition}
This type of equilibria can be categorised as Nash equilibria, as each player's strategy is optimal given the decisions of others.
It should be noted that, in general, the equilibria we consider differs from the typical market equilibrium in economics~\citep{arrow1954existence}, where supply equals demand.
There is a special case the types of equilibria are equivalent in binary prediction market with fair odds, where marginal utility maximisation clears the market~\citep{beygelzimer12Learning}.

The Stackelberg
equilibrium is achieved as long as the expected profit of the bookmaker is maximised since the Kelly betting rule \cref{eq:bettor-utility-function} has analytical solutions $\bet_t$ for each bettor. We note that such equilibrium is unique for common distributions of bettors' beliefs, see \cref{app: uniqueness-of-equilibrium} for more details.
\cref{lem:existance-of-maximiser} establishes the existence of a Stackelberg equilibria and \cref{lem:char-maximiser} shows that such equilibria will deviate from the bookmaker's belief \( g \). 
\negspace
\begin{restatable}{lemma}{existanceOfMaximiser}
    \label{lem:existance-of-maximiser}
    For any fixed $T$, the profit function $u_{1:T}$ in \cref{eq:argmax-profit} is upper-bounded, and it admits at least one maximiser $(\priceR^\star, \priceL^\star) \in (0,1)^2$.
\negspace
\end{restatable}
\negspace
\begin{restatable}{lemma}{charMaximiser}
    \label{lem:char-maximiser}
    Suppose the bettor belief distribution $f(x) > 0$ for all $x\in (0,1)$, then 
    for prices with non-zero overround $\priceR + \priceL >1$,
    all maximisers $(\priceR^\star, \priceL^\star)$ of profit satisfies
    \bignegspace
    \negspace
    \begin{equation*}
         1 - \priceL^\star < g < \priceR^\star.
    \bignegspace
    \negspace
    \end{equation*}
\end{restatable}
The combination of \cref{lem:existance-of-maximiser,lem:char-maximiser} clarifies the claim that the optimal prices, without fair odds, must necessarily deviate from the bookmaker's belief, 
which can be used 
to create a ``house edge'' in the prices.
\cref{prop:profit-margin-preference-deviation} presents a lower-bound of profit with the belief deviation between the bookmaker and the bettors.
\begin{restatable}{proposition}{profitMargin}
\label{prop:profit-margin-preference-deviation}
Let $u^{\star}$
denote the maximum utility corresponding to \cref{eq:argmax-profit}. Then
\(
u^{\star} \geq (g - \expect{\pt})^2.
\)
\negspace
\end{restatable}

\cref{prop:profit-margin-preference-deviation} and \cref{ssec:model_bookmaker} discussed the effect of mean bettor belief $\expect{\pt}$, the rest of this section establishes that a larger second-order deviation (or more diversity in bettor beliefs) will result in a larger profit.
We first define {\it second-order stochastic dominance} (SOSD), a well-established concept in economics, to formally describe bettor diversity.%
\begin{definition}[{\citet[Definition 2.1]{dentcheva2003optimization}}]
    Let $F_1$ and $F_2$ be {\cdf}s over the interval $(0,1)$. $F_1$ is SOSD over $F_2$ if and only if $S_1(Z) < S_2(Z)$ for all $Z\in (0,1)$ and $S_1(1) = S_2(1)$, where
    \bignegspace
    \negspace
    \begin{equation*}
        S_i(Z) = \int_0^Z F_i(z)\ \dmeas{}z, \quad i \in \{1,2\}.
    \bignegspace
    \negspace
    \end{equation*}
\end{definition}

The following shows that ordering belief distributions by SOSD implies the same order in the expected profits. %
\begin{restatable}{proposition}{StochasticDominance}
\label{lem:Stochastic-Dominance-Profit-Margin}
Fixing \( g \), let $u_1, u_2$ be profit functions \cref{eq:profit-cont} using preference distributions \( F_1, F_2 \), respectively. Assume the \pdf's $f_1, f_2$ satisfy $\supp(f_1) = \supp(f_2) = [0,1]$.
If \( F_1 \) is SOSD over \( F_2 \), then for any pair of prices \( \priceR \in (g, 1) \) and \( \priceL \in (1-g, 1) \), we have
\negspace
\negspace
\begin{equation*}
     u_1(\priceR, \priceL) < u_2(\priceR, \priceL); \quad u_1^\star < u_2^\star,
\bignegspace
\end{equation*}
where  $u_1^\star = \max_{\priceR,\priceL} u_1(\priceR, \priceL)$ and $u_2^\star = \max_{\priceR, \priceL}u_2(\priceR, \priceL)$.
\end{restatable}

\cref{lem:Stochastic-Dominance-Profit-Margin} matches the intuition that bookmakers can make more profit by exploiting stubborn bettors on both ends of the preference spectrum, \ie, when the variance is larger or when the distribution has heavier tails.

\input{figure/example}

\paragraph{Example: Profit maximisation and prediction aggregation are incompatible.}
We construct a family of belief distributions whose means are all $0.5$, each parameterised by $m$, $\Delta_1,\Delta_2$.
We assume the bookmaker's belief is $g=0.5$ too. We refer to $0.5$ as the ``common belief''.
The parameter $m = \expect{\pt \mid \pt\ge 0.5}$ lies between $0.5$ and $1$.
By design, $\expect{\pt \mid \pt< 0.5} = 1-m$.
When $m$ is larger, the distribution is more polarised.
The parameter $\Delta_1,\Delta_2$ lie between $0$ and $\min\{m-0.5,1-m\}$, which represent how spread the distribution is over the regions $\pt\ge 0.5$ and $\pt<0.5$ respectively.
The distribution has the following \pdf:
\bignegspace
\[
f_{m,\Delta_1,\Delta_2}(p) = \begin{cases}
\frac{1}{4\Delta_1}, & \text{if }p\in m\pm \Delta_1; \\
\frac{1}{4\Delta_2}, & \text{if }p\in 1-m\pm \Delta_2; \\
0, & \text{otherwise}.
\end{cases}
\bignegspace
\]
Under such belief distribution, the expected profit \cref{eq:profit-cont} admits a unique maximizer $(\priceR^\star,\priceL^\star)$. For several values of $m$, we compute the possible range of $\priceR^\star$ and $\priceL^\star$ when $\Delta_1,\Delta_2$ vary.
The result is summarised in \cref{table:range-of-astar}, with its derivation and explanation given in \cref{app: incompatible-profit-prediction}.

For all values of $m$ in the table, the range of $\priceR^\star$ and $\priceL^\star$ does not include the common belief.
Furthermore, as the polarisation parameter $m$ increases, $\priceR^\star,\priceL^\star$ can become as large as $0.70710$, which is 41.4\% more than the common belief.
The prices do not give useful indication of where the common belief locates.

When $\Delta_1,\Delta_2$ are different, we may have skewed prices.
In \cref{fig:profit-example}, we consider the case $m=0.75$, $\Delta_1 = 0.25$ and $\Delta_2 = 0.1$. The prices are $\priceR^\star = 0.70710$ and $\priceL^\star = 0.63395$, which have a relative difference of 11.5\%.
Consequently, even the prices after normalization, $(\frac{\priceR^\star}{\priceR^\star+\priceL^\star},\frac{\priceL^\star}{\priceR^\star+\priceL^\star})$,
do not give an accurate indication of the common belief.

With prices $(\priceR^*, \priceL^*)$ maximising profit, a natural question is ``how to find them''? 
The online setting sets the stage for an answer, however, there is a subtle (but important) difference in bettor and bookmaker behavior.
Bettors approaching the market at different times potentially face different prices updated by the bookmaker, where bettors behave in accordance to maximising their expected utility (in the Kelly sense) given the currently available price.
We present online algorithms to set the book in the next two sections.

%% file: figure/example.tex
\begin{figure}[!t]
\centering
\raisebox{-0.5\height}{\includegraphics[width=.17\textwidth,trim={10pt 0pt 40pt 0pt},clip]{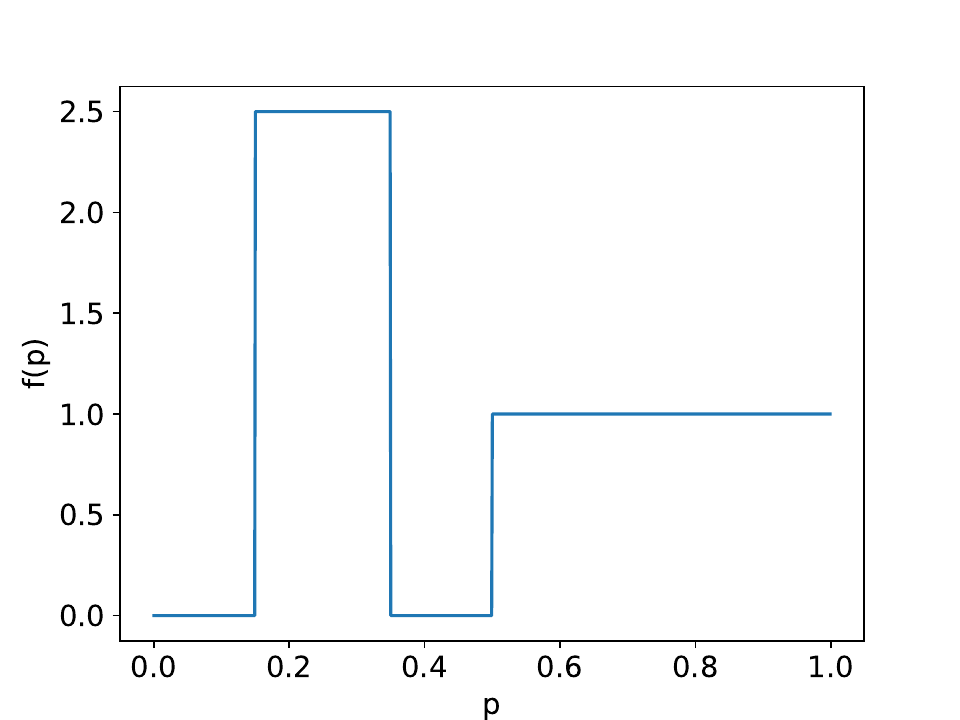}}
\raisebox{-0.5\height}{\includegraphics[width=.3\textwidth,trim={5pt 0pt 35pt 35pt},clip]{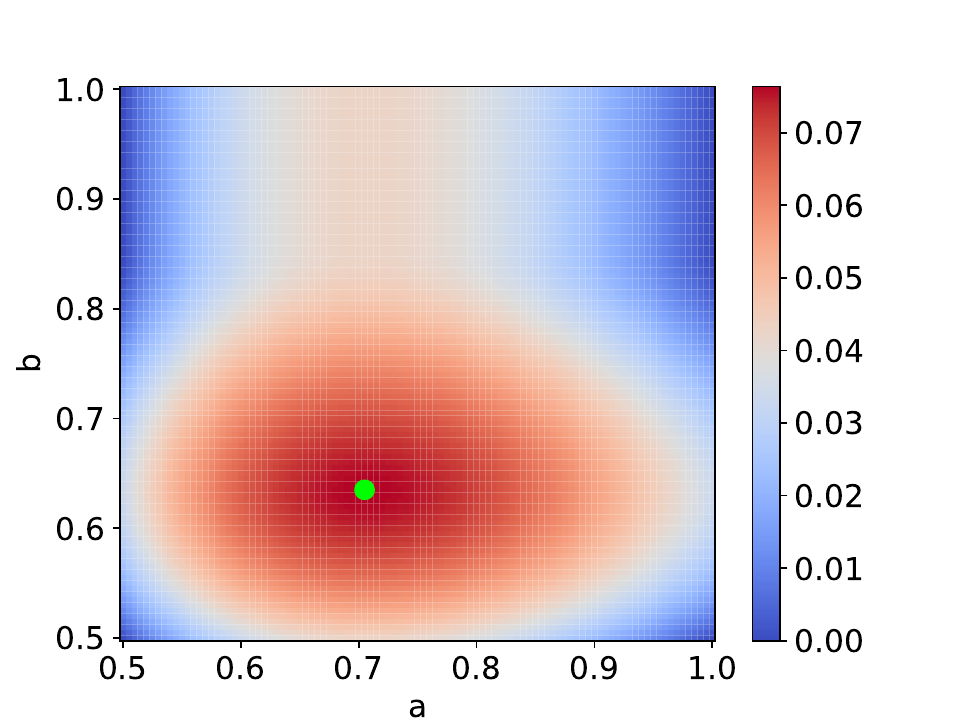}}%
\caption{\emph{Left}: Probability density function $f_{0.75,0.25,0.1}(p)$. \emph{Right}: Profit function, with profit-maximising prices $(\priceR^\star, \priceL^\star) = (0.70710,0.63395)$ marked by the {\color{olive}green point}.}\label{fig:profit-example}
\end{figure}
\begin{table}[t]
\begin{center}
\caption{\label{table:range-of-astar} Range of $(\priceR^\star,\priceL^\star)$ for different values of $m$.}
\begin{tabular}{ccc}
\toprule
$m$ & range of $\Delta_1,\Delta_2$ & range of $\priceR^\star,\priceL^\star$  \\
\midrule
$0.55$ & $(0~,~0.05]$ & $(0.52506~,~0.53353)$\\
$0.65$ & $(0~,~0.15]$ & $(0.57677~,~0.60608)$ \\
$0.75$ & $(0~,~0.25]$ & $(0.63395~,~0.70710)$ \\
$0.85$ & $(0~,~0.15]$ & $(0.70420~,~0.70710)$ \\
\bottomrule
\end{tabular}
\end{center}
\end{table}

%% file: sections/SA.tex
\section{Online Learning under Unfair Odds}
\label{sec:sa}

\paragraph{Setting.} 
We aim to find an optimal prices series $(\priceR_t,\priceL_t)$ in an online learning setting. 
Bettors $t \in \{1,\ldots,T\}$ arrive sequentially, each making a bet in response to the current prices \( (\priceR_t, \priceL_t) \). We assume that the bettors' belief distributions are \iid, and denote the \pdf and \cdf of their belief distributions as $f$ and $F$ respectively. The size of each bet \( \bet_t \geq 0 \) in dollars / wealth and the side taken (either \eventR or \eventL) are visible to bookmaker to update prices to \( (\priceR_{t+1}, \priceL_{t+1}) \). The bookmaker aims to maximise an {\it online} version of their own expected profit with the price series 
\begin{equation}
\label{eq:profit_online}%
u_{1:t} (\priceR_{1:t}, \priceL_{1:t}) \defeq \sum_{\tau=1}^t \ut(\priceR_\tau,\priceL_\tau)\cdot \expect{\wbettor_{\tau}}.
\bignegspace
\end{equation}
The bookmaker knows neither bettor belief \( \pt \) nor its distribution $f(p)$ a-priori, but it will gather information as the bets come in.
We assume that bookmaker has an estimate of the average wealth of bettors {\( \avgwealtht \approx \expect{\wt}\)}, which can be available via, \eg, the online wallet of bettor $t$ on a betting platform or the average behaviour among a set of bettors.
We estimate the belief of bettor $t$ by:
\begin{equation}
    \label{eq:est_belief}
    \widehat{\pt} =
    \begin{cases}
    \priceR_t + (1 - \priceR_t) \cdot \frac{\bet_t}{\avgwealtht},  & \textrm{if the bet is on \eventR};\\
    (1 - \priceL_{t}) \cdot \left(1 -\frac{\bet_t}{\avgwealtht}\right),  & \textrm{if the bet is on \eventL}.
    \end{cases}
\end{equation}
As the bettors are Kelly Bettors, by \cref{eq:kelly-betting-rule}, 
$\widehat{\pt}$ is an unbiased estimator of \( \expect{\pt} \) when \( \avgwealtht = \expect{\wt} \).
However, the estimates \( \widehat{\pt} \) may not necessarily lie in \( [0, 1] \).
To ensure that $\widehat{\pt}$ is a valid probability, we can either assume bet size $\bet_t \leq \avgwealtht $ (\cref{thm:SA-convergence}), or ``clip'' the value of $\widehat{\pt}$ to a subinterval of $[0, 1]$ as in the fair odds setting (\cref{sec:ftl}).

\paragraph{Stochastic Approximation (SA).}
The profit function \cref{eq:profit-cont} is non-concave in general, which presents a challenge in finding the optimal prices. 
One might want to consider utilising general first-order optimisation methods, \eg, gradient descent.
However, it is still challenging to directly apply these methods as the gradients $\nabla u_t(\priceR,\priceL)$ %
rely on the tail expectations, \eg, estimating 
the \cdf $F(\priceR)$ over a variety of prices \( \priceR \).
Nevertheless, the first-order optimality conditions of \( u_t(\priceR, \priceL) \) will become useful in deriving an algorithm for online profit maximisation.

\begin{restatable}{theorem}{FocGeneralCase}
    \label{lem:FOC-General Case}
    The first-order optimality condition could be reformulated as
    \begin{equation}
    \label{eq:SA-FOC}
    \left\{
    \begin{aligned}
        \Upsilon^{R}(\priceR) &\defeq G(\priceR) + \priceR -  \expect{\pt \mid \pt\geq \priceR} = 0;\\
        \Upsilon^{L}(\priceL) &\defeq G(\priceL) + \priceL -  \expect{\pt \mid \pt\leq 1- \priceL} = 0,
    \end{aligned}
    \right.
    \end{equation}
    where $G(x) = x(1-x)(x-\predict) / (x^2 - 2\predict x +\predict)$.
\end{restatable}

\cref{eq:SA-FOC} describes potential local maxima -- finding optimal prices is equivalent to a stochastic root-finding problem.
Similar problems have been extensively studied and can be efficiently solved via stochastic approximation (SA) algorithms~\citep{RM1951StochasticApproximation}.
\begin{definition}[{\citet{RM1951StochasticApproximation}}]
    \label{def:SA-process}
    For some twice differentiable function $h$, a stochastic approximation process is a process $(X_t)_{t\in \nn}$ adapted to the filtration $(\calF_t)_{t\in \nn}$ that admits the following form
    \[
    X_{t+1} =  X_t - \eta_{t+1}(h(X_t) + M_{t+1}),
    \bignegspace
    \]
    where $(M_{t})_{t\in\nn}$ is a random noise process satisfying
    \(
    \expect{M_{t+1} \mid \calF_t} = 0
    \),
    and the sequence of step sizes $(\eta_t)_{t\in \nn}$ satisfies
    $\sum_{t\in \nn} \eta_t = \infty$ and $\sum_{t\in \nn} \eta_t^2 < \infty$.
    In addition, $h(X_t)$ is uniformly bounded over $t\in \nn$.
\end{definition}

A special case of the SA algorithm is  when the function $h$ in \cref{def:SA-process} corresponds to an unbiased estimator of a function's gradient.
Without random noise, \ie \( M_{t} = 0 \) over all \( t \), the process becomes stochastic gradient descent (SGD)~\citep{RM1951StochasticApproximation,bottou2018optimization}. 
When the random noise is set to be sampled from a standard Gaussian, the process becomes stochastic gradient Langevin dynamics (SGLD)~\citep{borkar1999strong,welling2011bayesian}.
Both SGD and SGLD can be perceived as stochastic root-finding algorithms, aiming to locate points where an equation of ``gradient equals zero'' is satisfied. 

In this work, we aim to solve \cref{eq:SA-FOC} using SA algorithms, summarised in \cref{alg:SA}. In particular, to set prices we run two instances of SA where we set $h$ to be $\Upsilon^{R}(\priceR)$ and $\Upsilon^{L}(\priceL)$ as per \cref{lem:FOC-General Case}.
In addition the random noise $M_{t+1}$ are set to be the deviation between an estimate of the $t$-th bettor's belief and the true mean tail means of the bettor distribution, \ie, 
$M_{t+1} = \expect{\pt \mid \pt\geq \priceR_t} - \widehat{p_t}$ and 
$M_{t+1} = \expect{\qt \mid \qt\geq \priceL_t} - 1 + \widehat{p_t}$, respectively.
The resulting SA updates are given by \cref{eq:price_update_pos,eq:price_update_neg}.

\input{figure/sa}

\paragraph{Convergence.} Under mild assumptions, \cref{alg:SA} converges to a local maximum of the bookmaker's profit, which is formally stated in the following theorem.
\begin{restatable}{theorem}{saConvergence}
    Suppose that
    \bignegspace
    \begin{itemize} [itemsep=3pt,topsep=0pt,parsep=0pt]
       \item  The probability density function $\prefPDF$ is differentiable and the support of $\prefPDF$ satisfies $\supp(f) = [0,1]$;
       \item Bettors will not place bets exceeding the estimated wealth, \ie, $\bet_t \leq \avgwealtht$;
       \item The set of solutions to \cref{eq:SA-FOC} is finite;
       \item For $i\in \{L,R\}$ and for any $p$ satisfying $\Upsilon^i(p) = 0$, there exists a neighbourhood $\mathcal{N}$ of $p$ such that $\Upsilon^i(z)(z - p) < 0$ for all $z \in \mathcal{N} \setminus \{p\}$.

    \end{itemize}
    Then for sufficiently large $m, \gamma >0$, \cref{alg:SA} will almost surely converges to a local maximum of $\profit$ when using a learning rate $\eta_t = {\gamma}/{(t + m)}$.
    \label{thm:SA-convergence}
\end{restatable}
The second assumption in \cref{thm:SA-convergence} ensures us that the individual point-wise predictions of bettor's beliefs \( \bet_t / \avgwealtht \), for any \( t \), lies in the interval \( [0, 1] \). 
This further ensures $\widehat{\pt} \in [0, 1]$ as per \cref{eq:est_belief}.
For the last assumption, it is also sufficient to have the first derivatives of \( \Upsilon^R, \Upsilon^L \) (assuming differentiability) to be non-zero at the critical points.

The proof of \cref{thm:SA-convergence} can be divided into the following key steps. First, by exploiting the bettors' strategies, we demonstrate that the SA algorithm is well-defined according to \cref{def:SA-process}. Second, by identifying the explicit Lyapunov function of the process, we show that the SA dynamic almost surely converges to a critical point of $\profit$. 
Third, we adopt the saddle point avoidance results of \citet{pemantle1990nonconvergence} to rule out the possibility that the dynamic converges to critical points other than local maximisers.

Regarding the efficiency of the algorithm, we establish an $\mathcal{O}(1/\sqrt{T})$ convergence against the worst maximiser of $u_t$.

\begin{restatable}{theorem}{saRegretBound}
    \label{thm:SA-regret-bound}
    Suppose that the assumptions of \cref{thm:SA-convergence} holds.
    Let $(\priceR^\sharp,\priceL^\sharp)$ be the worst local maximiser of $u_t$,%
    \bignegspace
    \negspace
    \[
    (\priceR^\sharp,\priceL^\sharp) = \argmin_{(\priceR,\priceL)\in \calW} u_t(\priceR,\priceL),
    \bignegspace
    \negspace
    \]
    where $\calW$ is the set of all local maximisers of $u_t$. Then, there exists a finite constant $L_u > 0$ which is dependent on $u_t$, such that, for sufficiently large $T\in \nn$, we have
    \bignegspace
    \[
     \expect{u_T(\priceR_{T},\priceL_{T})}  \geq  u_t(\priceR^\sharp,\priceL^\sharp) - 7L_uT^{-1/2}. 
    \bignegspace
    \negspace
    \]
\end{restatable}
In particular, when the maximiser is unique, $(\priceR^\sharp, \priceL^\sharp) = (\priceR^\star, \priceL^\star)$ is the global maximiser and hence the equilibrium price. 
In our analysis, we note that the process will converge to one of the local maximisers, which is guaranteed to have a neighbourhood where $u_t$ is locally Lipschitz smooth and concave. By the fact that the process will be in one of such neighbourhoods when $t$ is large enough, we could characterise the convergence rate under such assumptions. The detailed analysis is given in \cref{app: proof-regret-bound}.
Moreover, by conducting the non-escape analysis, similar to~\citet[Theorem 4]{mertikopoulos2020almost}, we obtain a stronger result which clarifies which local maximiser the process converges to.
\begin{restatable}{theorem}{saRegretConditioned}
    \label{thm: SA-regret-conditioned}
    Suppose that the assumptions of \cref{thm:SA-convergence} holds.
    For any constant $\delta<\frac{1}{2}$, let $(\priceR^\flat,\priceL^\flat)$ be one of the local maximisers of $u_t$. Then there exist neighbourhoods $\mathcal{U}_1$ and $\mathcal{U}$ of $(\priceR^\flat,\priceL^\flat)$ such that, if $(\priceR_{1},\priceL_{1})\in \mathcal{U}_1$, the event
    \[
    \Omega_\mathcal{U} = \{(\priceR_{t},\priceL_{t})\in \mathcal{U}~\text{for all}~t\in\nn\}
    \]
    occurs with probability at least $1-\delta$. Moreover, there exists a finite constant $L_u > 0$ which is dependent on $u_t$, such that, for sufficiently large $T\in \nn$, we have
    \[
    \expect{\left\vert u_t(\priceR_{T}^\flat,\priceL_{T}^\flat) - u_t(\priceR_{T},\priceL_{T}) \right\vert  ~\mid~ \Omega_\mathcal{U}}\leq 4\sqrt{6}L_u T^{-1/2}.
    \bignegspace
    \]
\end{restatable}
Our proof mainly follows \citet{mertikopoulos2020almost} but addresses several additional technical difficulties. For example, our SA algorithm does not perform the exact gradient step and every bettor does not cause prices to be updated (they however will contribute to the total time complexity).

\bignegspace
\bignegspace
\paragraph{Regret.} Although the previous results characterise the limiting case when \( T \) is large, we also want to characterise the cumulative penalty across all timesteps/bettors \( t \in \{1 \ldots T\}\). This can be concretely described by \emph{regret}~\citep{hazan2014beyond,hazan2016introduction}. In the case of our bettors and bookmaker profits, we define stochastic regret in the following.

\begin{definition}
    \label{rmk: Regret}
    Given a sequence of bookmaker prices \( (\priceR_1, \priceL_1), \ldots, (\priceR_T, \priceL_T) \), the stochastic regret against $(\priceR, \priceL)$ is
    \negspace
    \begin{equation}
        \label{eq:regret}
        \regret(T,\priceR,\priceL) \defeq u_{1:T}(\priceR, \priceL)  -  u_{1:T}(\priceR_{1:T}, \priceL_{1:T}).
    \bignegspace
     \negspace
     \negspace
    \end{equation}
\end{definition}
Standard stochastic regret~\citep[Section 2.2]{hazan2014beyond} is given by \( \regret(T, \priceR^{\star}, \priceL^{\star}) \) with \( (\priceR^{\star}, \priceL^{\star}) = \argmax_{\priceR, \priceL} u_t(\priceR, \priceL) \).
Since $u_t(\priceR_t,\priceL_t)$ is bounded %
for all $t\in \nn$, the following result on {\it local} stochastic regret can be obtained immediately.
\begin{corollary}
\label{cor:sa_regret}
    Suppose that the assumptions of \cref{thm:SA-convergence} holds.
    Then we have
    \[
     \expect{\regret(T,\priceR^\flat,\priceL^\flat) \mid  \Omega_{\mathcal{U}}} = \bigoh(\sqrt{T}).
     \bignegspace
     \negspace
    \]
\end{corollary}
The expected regret without conditioning can also be derived as $\mathcal{O}(\sqrt{T})$, as per \cref{thm: SA-regret-conditioned} and its assumptions.

%% file: figure/sa.tex
\begin{algorithm}[tb]
\caption{Online SA Algorithm}\label{alg:SA}%
\begin{algorithmic}[1]
\REQUIRE Wealth estimate $(\avgwealtht)_{t\in \nn}$, ground-truth estimate $g$, initial price $(\priceR_1, \priceL_1)$.
\FOR{each bettor entering the market at time $t$}
   \STATE Receive the bet placed by the bettor $\bet_t$.
   \STATE Estimate bettor belief $\widehat{\pt}$ with $(\priceR_t, \priceL_t)$ via \cref{eq:est_belief}.
   \IF{bettor \( t \) bets for \eventR}
       \STATE Update the price $\priceR_{ t+1}$ as follows:%
       \par
       \parbox{\linewidth}{%
       \vspace{-12pt}
       \begin{align}
           \label{eq:price_update_pos}
           \priceR_{t+1} =  \priceR_{t} - \eta_{t+1}( \priceR_{t} + G(\priceR_{t})- \widehat{\pt}).
       \end{align}
       }\par
       \vspace{-5pt}
   \ELSIF{bettor \( t \) bets for \eventL}
       \STATE Update the price $\priceL_{ t+1}$ as follows:
       \par
       \parbox{\linewidth}{%
       \vspace{-12pt}
       \begin{align}
           \priceL_{t+1} =  \priceL_{t} - \eta_{t+1}( \priceL_{t} + G(\priceL_{t})- 1 + \widehat{\pt} ).
           \label{eq:price_update_neg}
       \end{align}
       }\par
       \vspace{-5pt}
   \ENDIF
\ENDFOR
\end{algorithmic}
\end{algorithm}

%% file: sections/FTL.tex
\bignegspace
\bignegspace
\section{Online Learning under Fair Odds}
\label{sec:ftl}
\negspace
\negspace

\input{figure/ftl}

In this section, we assume that the bookmaker maintains fair odds \( \priceR + \priceL = 1 \). 
We note that \cref{alg:SA} cannot be directly used for the fair odds case due to two reasons. 
.
First, prices (and their optimal value) must be restricted to a region \( (\priceR_t, \priceL_t) \in (g, 1) \times (1-g, 1) \) in \cref{thm:SA-convergence}. 
Second, the convergence of \cref{thm:SA-convergence} relies on the prices \( (\priceR_t, \priceL_t) \) staying within a region where profit \( u_t(\priceR_t, \priceL_t) \) is lower-bounded.
We do not have the same control to restrict \( \priceR \) under fair odds from reaching the boundaries of \( [0, 1] \), where the gradient of the profit \( \priceR \mapsto u_t(\priceR, 1-\priceR)\) can be infinite.
Instead, we exploit the overall concavity of the profit function to adapt a \emph{follow the leader} (FTL)~\citep{hazan2016introduction} algorithm for fair odds, as presented in \cref{alg: FTL}.
Although our FTL algorithm has worse asymptotic regret bounds than \cref{alg:SA}, it will have better convergence conditions -- importantly, FTL's convergence is to the global maxima unlike \cref{alg:SA}.

\paragraph{Follow The Leader (FTL).}
As we have fair odds, \( \priceL_t = 1 - \priceR_t \), our goal is to maximise profit via the map $\priceR \mapsto u_t(\priceR, 1 - \priceR)$, which is concave, see \cref{eq:fair_profit}. Thus, by taking the first-order optimality condition of $u_t(\priceR, 1 - \priceR)$, we get the following closed-form expression of the (unique) maxima \( \priceR^\star = \psi(\expect{\pt}) \), where
\bignegspace
\negspace
\begin{equation}
    \psi(p)
     \defeq \frac{\sqrt{gp}}{\sqrt{gp} + \sqrt{(1-g)(1-p)}}. \label{eq: opt-price-fair}
\bignegspace
\end{equation}
Notably, the optimal price \cref{eq: opt-price-fair} does not require knowledge of the entire belief distribution \( f \) -- only the first moment \( \expect{\pt} \) is needed. As a result, to find the optimal price \( \priceR^{\star} \) in an online setting, we only need to obtain an online estimate of the expected bettor's belief \( \expect{\pt} \).

To estimate the \( \expect{\pt} \), we utilise a cumulative average -- taking into account all prior bets \( t' < t \) -- which is equivalent to making the optimal estimate of \( \priceR_t^\star = \psi(\expect{\pt}) \) in hindsight, \ie, a FTL algorithm~\citep{hazan2016introduction}.
To ensure that our cumulative average of bettor belief does not fall outside of \( [0, 1] \), we clip each \( \widehat{\pt} \) to a region \( [\tau, 1-\tau] \) for \( \tau \in (0, 0.5) \).

\input{figure/two_gau}

\paragraph{Convergence.}
We are interested in analysing how the learned prices \( (\priceR_{T}, \priceL_{T}) \) learned by \cref{alg: FTL} converges to the optimal prices \( (\priceR^{\star}, \priceL^{\star}) \) \wrt utility \( u\).
A stochastic convergence result for \cref{alg: FTL} can be obtained by considering the local Lipschitz properties of functions.

\begin{restatable}{theorem}{regretBoundFTL}
    \label{thm:regret-bound-FTL}
    Suppose that both $g$ and $\expect{p_t}$ lie in the open interval $(\tau,1-\tau)$
    and $w_t$ is uniformly bounded above by an absolute constant, almost surely.
    Suppose \( (\priceR_t, \priceL_t)_{t\in\mathbb{N}} \) is the price sequence generated by \cref{alg: FTL}.
    Then there exists a finite \( L > 0\)  such that for any \( \delta > 0 \) and for sufficiently large $T \in \mathbb{N}$, we have
    \bignegspace
    \negspace
    \begin{equation*}
        u_t(\priceR_{T}, \priceL_{T}) \geq  u_t(\priceR^\star, \priceL^\star) - L T^{-1/2} \sqrt{\log(1/\delta) },
    \negspace
    \bignegspace
    \end{equation*}
    
    with probability at least \( 1- \delta \).
\end{restatable}
The constant \( L \) in \cref{thm:regret-bound-FTL} corresponds to a product local Lipschitz constants of \( u \) and \( \psi \) when restricting \( \overline{\pt} \) to \( (\tau, 1-\tau) \). In general, a larger $\tau$ will result in a larger $L$.

\paragraph{Regret.} \cref{thm:regret-bound-FTL} can be restated \wrt a high probability regret bound~\citep{bubeck2012regret,bartlett2008high}.
\begin{corollary}
    \label{cor:ftl_regret}
    Suppose that the assumptions of \cref{thm:regret-bound-FTL} holds.
    Then with probability \( 1 - \delta\), we have
    \bignegspace
    \negspace
    \begin{equation*}
        \regret(T, \priceR^{\star}, \priceL^{\star}) = \bigoh(\sqrt{T \log T}).
        \negspace
        \bignegspace
    \end{equation*}
\end{corollary}
The proof is immediate from a union bound, yielding sublinear regret.
Although the regret when compared to 
\cref{alg:SA}
is worse by a factor of \( \bigoh(\sqrt{\log T}) \) (\cref{cor:sa_regret}) there are notable differences which make the FTL bound preferable.
 For instance, in \cref{thm:regret-bound-FTL} convergence is \wrt to the global optimal prices \( (\priceR^\star, \priceL^\star) \) whilst \cref{thm: SA-regret-conditioned} is only \wrt a local maxima. Furthermore, in \cref{thm:regret-bound-FTL} we do not need to condition our price sequence on \( \Omega_{\mathcal{U}} \) to achieve a convergence rate.

As the fair odds setting of betting markets coincides with prediction markets when \( g = \expect{\pt} \), one may want to consider FTL's regret in comparison to prediction markets.
From \citet[Corollary 1]{Frongillo:2012}, we know that prediction market dynamics follow online mirror descent updates, where the maximised function is a KL-divergence \( \priceR \mapsto - \kl(\expect{\pt} 
\Mid \priceR) \). The prediction market's regret~\citep{duchi2010composite} \wrt this KL-utility can be shown to be \( \bigoh(\sqrt{T \log T})\) -- taking appropriate step sizes~\citep[Chapter 5.3]{hazan2016introduction} and a union bound over \( \delta / T \) steps, similar to \cref{cor:ftl_regret} -- matching the FTL regret. As such, despite the difference in utilities being maximised, our regret matches automated market makers derived from strictly convex functions~\citep{abernethy2013efficient}.

%% file: figure/ftl.tex
\begin{algorithm}[tb]
\caption{Follow The Leader}\label{alg: FTL}
\begin{algorithmic}[1]
\REQUIRE Wealth estimate $(\avgwealtht)_{t\in \nn}$, ground-truth estimate $g$, clipping value \( \tau \in (0, 0.5) \), initial price $\priceR_1 $.
\FOR{each bettor entering the market at time $t$}
   \STATE Receive the bet placed by the bettor $\bet_t$.
   \STATE Estimate bettor belief $\widehat{\pt}$ with $(\priceR_t, 1-\priceR_t)$ via \cref{eq:est_belief} %
   \STATE Update the estimate of expected belief as
   \par
   \parbox{\linewidth}{%
   \vspace{-6pt}
   \begin{align}
        \overline{\pt} &= \frac{t-1}{t} \cdot \overline{\pbettor}_{t-1} + \frac{1}{t} \cdot 
        \widehat{\pt}.
        \label{eq:cumavg_belief_noclip}
   \end{align}  
   }\par
   \vspace{-3pt}
   \STATE Following \cref{eq: opt-price-fair}, update the price with a clipped cumulative average:
   \par
   \parbox{\linewidth}{%
   \vspace{-3pt}
   \begin{equation*}
       \priceR_{t+1} = \psi\left(\clip(\overline{\pt}; \tau, 1-\tau)\right).
   \end{equation*}
   }\par
\ENDFOR \hfill // \( \clip(x; l, u) \defeq ((x \vee l) \wedge u) \) for $ l < u $
\end{algorithmic}
\end{algorithm}

%% file: figure/two_gau.tex
\begin{figure*}[ht!]
    \centering
    \negspace
    \includegraphics[width=1.\textwidth]{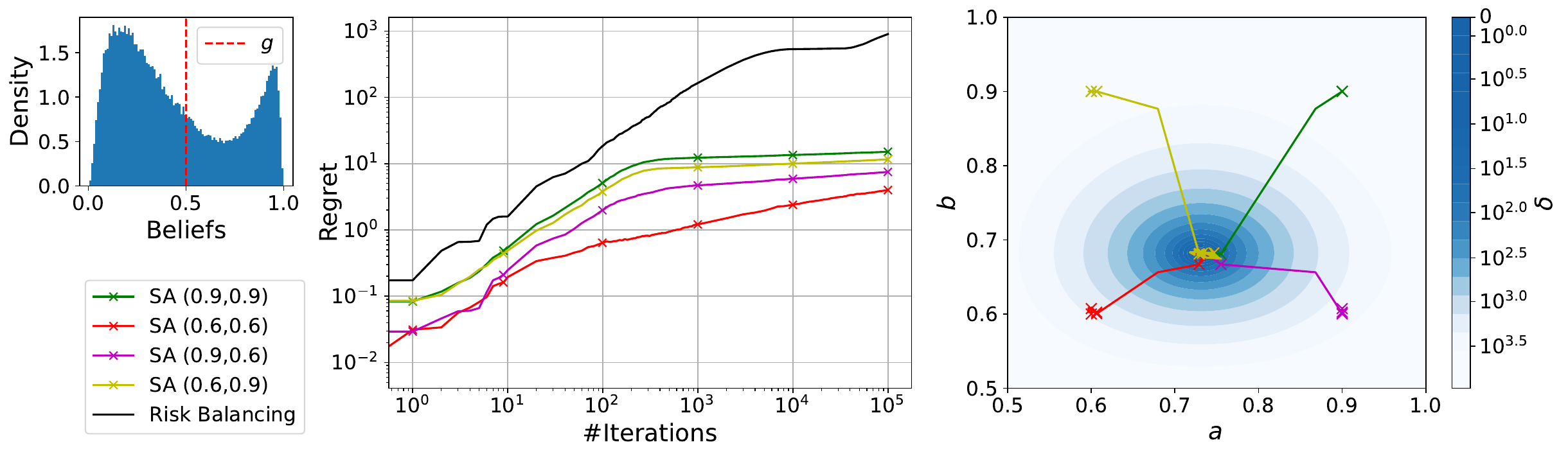}%
    \bignegspace
    \bignegspace
    \caption{Simulation of \Cref{alg:SA} with 100,000 bettors. \emph{Left}: The distribution of bettors' beliefs, unknown to bookmaker. \emph{Middle}: Regrets over 100,000 iterations, comparing \Cref{alg:SA} under different initialisation to risk-balancing. Markers ``$\times$" indicate number of iterations on a log-scale $\{10^1, \ldots,10^5\}$. \emph{Right}: Contour plot of $\delta(a,b) \defeq  u_{1:T}(\priceR^\star,\priceL^\star) - u_{1:T}(\priceR,\priceL)$ where \( T = 10^5 \), darker colours means closer to maximum profit.
    }\label{fig:two-gau}%
\bignegspace
\bignegspace
\bignegspace
\end{figure*}

%% file: sections/Empirical.tex
\bignegspace
\bignegspace
\section{Empirical Results}
\label{sec:Empirical}
\bignegspace
\bignegspace
We 
illustrate the efficiency of \cref{alg:SA,alg: FTL} empirically\footnote{Code and data to reproduce results are found at: \url{https://github.com/haiqingzhu543/Betting-Market-Simulation-2024}.}.
An advantage of our theoretic results is that they hold for a wide range of bettor belief distributions, only requiring weak assumptions. Our empirical analysis aims to elucidate how different properties of the belief distributions (not captured by theory) change the performance of our algorithms.
\bignegspace
\bignegspace
\bignegspace
\paragraph{Settings.} We set the bookmaker's belief $g = 0.5$ throughout all simulations. We generate $10^5$ Kelly bettors with a mixture of beliefs -- one Gaussian for event \eventR and \eventL respectively, followed by a sigmoid function to ensure that beliefs lie within (0, 1), \ie $\pt = \text{sigmoid}(s_t),~{t = 1,\ldots,10^5}$ 
with $s_t \sim 0.25 \cdot \mathcal{N}(2,1) + 0.75 \cdot \mathcal{N}(-1,1)$. 
The histogram of the distribution is in \cref{fig:two-gau} (Left). We note that the distribution has one mode on either side of $g$ but is not symmetric around $g$. 
We compute regret using \cref{rmk: Regret}, where the optimal price is$(\priceR^\star, \priceL^\star) = \argmax_{\priceR + \priceL \geq 1} u_t(\priceR, \priceL)$ generally, and under fair odds $\priceR^\star = \argmax_{\priceR \in (0,1)}u_t(\priceR, 1- \priceR)$. 

\cref{fig:two-gau} summarises our observations of \cref{alg:SA}. We use four different initialisations, and set the learning rate as $\eta_{t+1} = 300/(t+5000)$. As a baseline, we compare this to a {\it risk-balancing} heuristic~\cite{levitt2004gambling} where bookmakers try to equalise the number of dollars wagered on each outcome, the implementation is described in \cref{app-alg: RB}. 
\cref{fig:two-gau} (Middle) shows that  under all initialisations, the SA algorithm could maintain low regret $\leq 10^2$. However, the regret under the risk-balancing scheme is larger by more than an order of magnitude, and keeps increasing (note y-axis is in log scale). 
\cref{fig:two-gau} (Right) shows the trajectories of the price dynamics - all trajectories converge to the global maximiser. Further, when $t\geq 10^4$, the trajectories stay within the last contour, verifying %
\cref{thm:SA-convergence,thm: SA-regret-conditioned}, \ie the dynamic will converge to a (local) maximiser and stays in the neighbourhood of the maximiser if it enters it. 
Overall, when the maximiser is unique, \cref{alg:SA} is robust across different initialisations, converges fast and suffers low regret. %
If the \pdf has multiple modes on each side, it will result in different landscapes of the profit maximisation problems, one such case is in \cref{app: Empirical}.

\input{figure/fair}

\cref{fig:Fair} examines \cref{alg: FTL} (FTL) empirically. We compare it to the market-making approach via the Logarithmic Market Scoring Rule (LMSR) for Kelly bettors \cite{hanson2007logarithmic, abernethy2013efficient,beygelzimer12Learning} which is primarily designed for prediction markets to elicit the bettors' beliefs. \cref{fig:Fair} (Left) presents the regrets over time for both algorithms. It is observed that the regret of FTL stays $\leq 10^1$ throughout and grows slowly, validating \cref{thm:regret-bound-FTL}. On the other hand, the regret of LMSR grows faster than FTL. \cref{fig:Fair} (Right) presents the trajectories of $\priceR_t$ of both algorithms. Both processes converge to steady-states quickly but there is a noticeable gap between the limits. We found that the price dynamic of LMSR approximately converges to the average belief of the crowd. As expected, FTL converges to a point between the bookmaker's belief $g =0.5$ and the crowd's average belief. Such findings echo the insight that the bookmaker exploits 
the bias of bettors to maximise the profit, \cf discussion in \cref{ssec:model_bookmaker}.

%% file: figure/fair.tex
\begin{figure}[h!]
\begin{minipage}{0.24\textwidth}
  \centering
  \includegraphics[width=1.\textwidth]{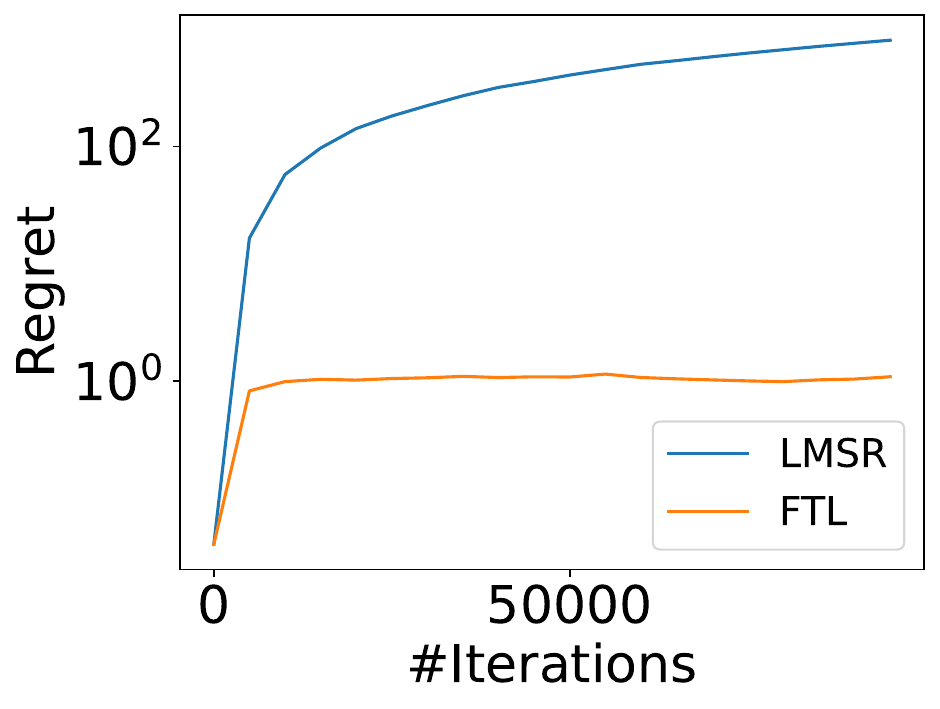}
\end{minipage}%
\begin{minipage}{0.24\textwidth}
  \centering
  \includegraphics[width=1.\textwidth]{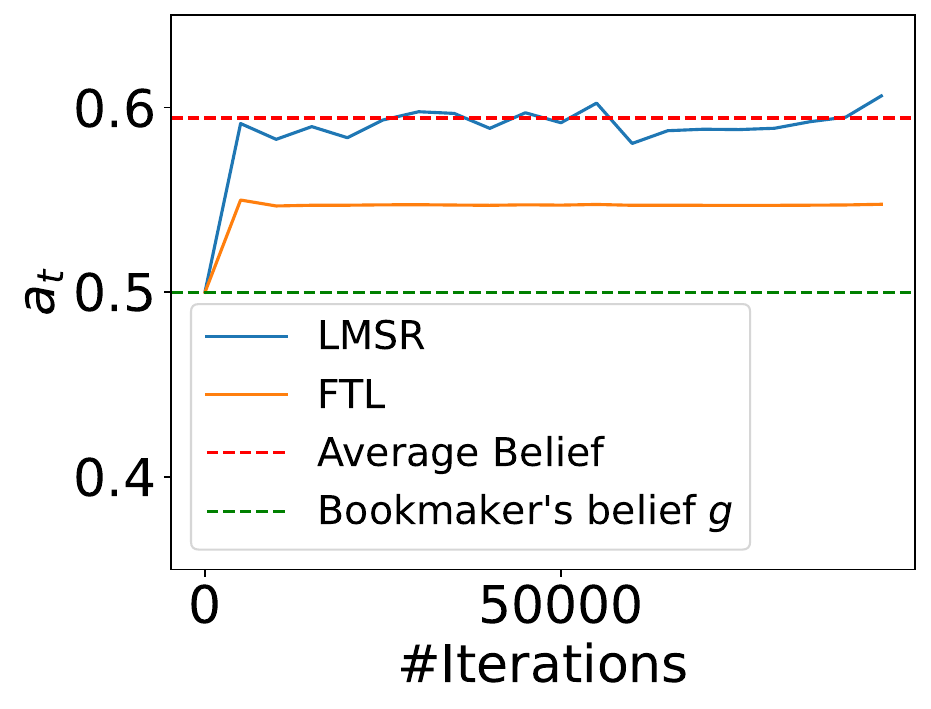}
\end{minipage}%
\bignegspace
    \caption{Simulating FTL and LMSR over 100,000 iterations. \emph{Left}: Regret. \emph{Right}: Price trajectory $\priceR_t$.}\label{fig:Fair}
    \bignegspace
    \bignegspace
    \bignegspace
\end{figure}

%% file: sections/Conclusion.tex
\bignegspace
\bignegspace
\section{Conclusion}
\bignegspace
\negspace

We articulate a binary
betting market model among Kelly bettors and utility-maximising bookmakers. 
This model encompasses prediction markets and betting markets with fair and unfair odds. It pinpoints the conflict between profit-making and eliciting predictions from the crowd -- a fact known by others \cite{ChenPennock2007}, but 
not explicitly connected to betting markets.
This model provides rigorous justifications of the empirical observations on bookmakers exploiting bettor bias for profit-making \cite{levitt2004gambling} -- answering an old academic joke ``we know it works in practice, but does it work in theory?” \cite{wolfers2006interpreting}.
Connecting these two insights motivated us to introduce modern online learning methods into bookmaker strategy, proposing two algorithms for finding optimal prices. While online learning strategies abound for prediction markets \cite{ChenV2010,Frongillo:2012}, ours might be the first implementable for a betting market. 

Directions for future work include:
(i) extensions to non-binary markets;
(ii) extensions to alternative bettor behaviours, including non-Kelly bettors, non-\iid beliefs, and belief distribution supported by a strict subset of interval $[0,1]$;
(iii) trade-offs between prediction and profit-making;
(iv) quantification of players' power in betting markets; and %
(v) conditions for unique equilibrium.

%% file: sections/ack.tex
\section*{Acknowledgements}
This work is supported in part by the Australian Research Council (ARC) projects DP240100506 and FT230100563.
We thank Rui Zhang and {\'{A}}lvaro Flores for their insightful comments on earlier drafts of this paper.
We thank Yanjun Liu for pointers to the theory of stochastic approximation.

%% file: sections/ImpactStatement.tex
\bignegspace
\negspace
\section*{Impact Statement}
\negspace
\negspace

This work presents several theoretical advances on binary betting markets. 
Betting (\eg sports betting and many other varieties) is a domain loaded with ethical and societal concerns. We will discuss the potential benefits and harms on several relevant aspects on betting, and then on potential impacts in other domains. 

{\it Theoretical understanding on bookmaker behaviour as public knowledge} is relatively scant compared to those on bettor behaviour and betting strategies in both research~\cite{thorp1975portfolio} and popular culture\footnote{\url{https://en.wikipedia.org/wiki/21_(2008_film)}}. Much knowledge on bookmaker operation are hidden under commercial confidentiality, available literature are either empirical~\cite{levitt2004gambling} or whistle-blowing on a particular aspect\footnote{\url{https://www.theguardian.com/society/2022/feb/19/stake-factoring-how-bookies-clamp-down-on-successful-gamblers}}. Therefore we deem new theoretical understanding on what makes bookmaker profitable, and how much the profit could be to have significant potential benefits. On the value of giving bookmakers a strategy to update prices frequently, a potential benefit is to have a public strategy that could level the field of gambling industry. A potential harm could be to increase or decrease revenue for specific bookmakers, thereby affecting fairness, or negatively affect a firm or its bettor population.

{\it Bettor behaviour} is deemed to follow a particular model \cite{kelly1956new} in this work. Our study focuses on bookmaker behaviour, but we do not rule out potential flow-on effects that influence the bettors and could cause potential harm. This could include peripheral measures that bookmakers adopt (other than adjusting odds and prices) to limit or encourage bettors to bet. 

{\it The gap between theory and practice} is non-trivial for this work. First of all, most betting markets are non-binary (including point spread in sports, or overlapping events), and the price structure is more complex than what is assumed in this work. Even if someone wants to adopt the online learning strategies in pricing, it might help or hurt the overall profit in a complex operations environment influenced by many other factors.

Finally, our eventual goal is to translate the methods and insights from binary betting to large-scale online behaviour mediated by algorithms.
Drawing on the analogy with multi-armed bandit problems being applicable to domains as diverse as clinical trials to adaptive routing, it is conceivable that bookmaking strategy can be applicable to domains that either elicit or exploit belief of the crowd. In particular, this work quantifies {\it platform power} represented by bookmaker profit obtained using prices as the instrument. We hope the methodology could help understand other large-scale online platforms such as social media and online attention (potential benefit), but also acknowledge that giving online platforms an implementable profit maximisation scheme may further increase platform power (potential harm to societal values).

%% file: appendix.tex
\appendix
\onecolumn

\counterwithin{theorem}{section}
\setcounter{figure}{0}
\setcounter{table}{0}

\renewcommand\thesection{\Alph{section}}
\renewcommand\thesubsection{\thesection.\Roman{subsection}}
\renewcommand\thesubsubsection{\thesection.\Roman{subsection}.\arabic{subsubsection}}

\renewcommand{\thetable}{\Roman{table}}
\renewcommand{\thefigure}{\Roman{figure}}

\newcommand{\tocrow}[1]{%
    \noindent $\hookrightarrow$ \cref{#1}: \nameref{#1} \hrulefill Pg \pageref{#1}\\
}

\makeatletter
\newcommand\setcurrentname[1]{\def\@currentlabelname{#1}}
\makeatother
\begin{center}
\Huge{Supplementary Material}
\end{center}
\begin{center}
This is the Supplementary Material to Paper "\papertitle". To
differentiate with the numberings in the main file, the numbering of Theorems is letter-based (A, B, ...).
\end{center}
\section*{Table of contents}
\noindent \textbf{Additional discussions}\\
\tocrow{app:notation}
\tocrow{sec:app_cvar}
\tocrow{app:imprecise_belief}
\tocrow{app: uniqueness-of-equilibrium}
\tocrow{app: incompatible-profit-prediction}
\tocrow{app: Empirical}

\noindent \textbf{Proofs and derivations}\\
\tocrow{appendix_kelly}
\tocrow{app: derivation of profit}
\tocrow{app: missing-proof-sec3}
\tocrow{app: FOC-General-case}
\tocrow{app: SA-convergence proof}
\tocrow{app: proof-regret-bound}
\tocrow{app: Proof-SA-regret-conditioned}
\tocrow{app: regret-bound-online-algo-fair}

\vfill

\newpage
\input{appendix/notation-table}
\input{appendix/cvar}
\input{appendix/imprecise-belief}
\input{appendix/Discussion-uniqueness}
\input{appendix/Incompatible}

\input{appendix/additional-empirical}

\input{appendix/Proof-for-section-2}
\input{appendix/Profit-Margin-Proofs}

\input{appendix/online-algorithm-general}
\input{appendix/online-algorithm-fair}

\vfill

%% file: appendix/notation-table.tex
\section{Table of Notations}
\setcurrentname{Table of Notations}
\label{app:notation}

\begin{table}[h]
\centering
\begin{tabular}{lll}
\toprule
\textbf{Symbol} & \textbf{Meaning}                         & \textbf{Defined} \\ 
\midrule
$\nn$ & Natural numbers $\{1,2,\ldots\}$\\
$x \wedge y$ & $\min\{x, y\}$  \\
$x \vee y$ & $\max\{x, y\}$  \\
$(x)_+$ & $\max\{x, 0\} = x \vee 0$  \\
$\clip(x; l, u)$ & $ \min\{ \max\{ x, l \}, u \} = ((x \vee l) \wedge u)$ for $l < u$ & \cref{alg: FTL} \\
\eventR, \eventL & Outcomes of binary events (\eventR = $\neg$ \eventL) & \Cref{sec:model}                \\
$t$             & Timestamp and index for bettors, assuming arriving sequentially                              & \Cref{sec:model}                \\ 
$T$ & Total number of bettors considered &  \Cref{sec:equilibria}                 \\ 
$\priceR$       & Price for event \eventR, $0<\priceR<1$                                        & \Cref{sec:model}                \\ 
$\priceL$       & Price for event \eventL, $0<\priceL<1$                                        & \Cref{sec:model}                \\
$g$       & Bookmaker's belief                                      & \Cref{sec:model}                \\ 

$\pt$           & Bettor $t$'s belief in event \eventR                                        & \Cref{ssec:model_bettor}                  \\ 
$\qt$               & Bettor $t$'s belief in event \eventL                                       & \Cref{ssec:model_bettor}                  \\ 
$\wt$               & Bettor $t$'s wealth  &    \Cref{ssec:model_bettor}           \\ 
$\bet_t$               & Absolute value of bettor $t$'s investment (bet)                                        & \Cref{ssec:model_bettor}                  \\ 
$f$             & \pdf of bettors' belief distribution                                & \Cref{sec:sa}                \\ 
$F$             & \cdf of bettors' belief distribution                                & \Cref{sec:sa}                \\ 
$(\priceR^\star, \priceL^\star)$               & Optimal prices (equilibrium prices)                                       &  \Cref{def: stackelburg-equilibrium}              \\ 
$(\priceR^\sharp,\priceL^\sharp)$ & Worst local maximiser of $u_t$                                     &  \Cref{thm:SA-regret-bound}              \\ 
$(\priceR^\flat,\priceL^\flat)$  & One of the local maximisers of $u_t$                                     &  \Cref{thm: SA-regret-conditioned}              \\

$\varphi_t^{\priceR} $             & Expected logarithm of wealth after bettor $t$ bets on event \eventR                                         & \Cref{ssec:model_bettor}                \\ 
$\varphi_t^{\priceL}$               & Expected logarithm of wealth after bettor $t$ bets on event \eventL                                        & \Cref{ssec:model_bettor}                 \\ 
$\ut(\priceR, \priceL)$               & The bookmaker's expected profit at time $t$                                        & \Cref{ssec:model_bookmaker}                \\ 
$u(\priceR, \priceL)$               & Same as $\ut(\priceR, \priceL)$; used in the appendix only 

&     \\ 
$u_t(\priceR, 1-\priceR)$               & The bookmaker's expected profit at time $t$ when the odds are fair                                         &  \Cref{ssec:model_bookmaker}                 \\ 
$u_{1:T}(\priceR,\priceL)$               & Bookmaker's cumulative profit                                        &  \Cref{sec:equilibria}                 \\ 

$u_{1:t} (\priceR_{1:t}, \priceL_{1:t})$               &  Online version of their own expected profit                                         &  \Cref{sec:sa}                 \\

$\avgwealtht$   & Bookmaker's estimate of the average wealth of bettor $t$                                         &  \Cref{sec:sa}                 \\ 
$\widehat{\pt}$ & Bookmaker's estimate the belief of bettor $t$                                     &  \Cref{sec:sa}                 \\ 
$\eta_t$ & Learning rate of the \Cref{alg:SA}                                     &  \Cref{alg:SA}              \\ 
$\Upsilon^{R}$, $\Upsilon^{L}, G$ & Auxiliary functions for characterising the first-order condition                                     &  \Cref{lem:FOC-General Case}                 \\ 
$\regret(T,\priceR,\priceL)$& Regret of online algorithms                                    &  \Cref{rmk: Regret} \\

\bottomrule
\end{tabular}
\end{table}

%% file: appendix/cvar.tex
\section{Connecting \( \cvar \) and Expected Profit}
\setcurrentname{Connecting \( \cvar \) and Expected Profit}
\label{sec:app_cvar}

    One interesting perspective of \cref{eq:profit-cont} is its connection to conditional value-at-risk ($\cvar$)~\citep{rockafellar2000optimization}.
    Suppose that both terms in \cref{eq:profit-cont} are positive, and that \( \priceR \) and \( \priceL \) are sufficiently far from \(g\).
    We derive a lower bound of $u_t(a,b)$ in terms of $\cvar$ values of the belief distribution. Thus, one can conclude that the utility in \cref{eq:profit-cont} is lower bounded by the tail behaviour of \( \pt, \qt \). For any random variable $\mathsf{X}$ of the belief distribution, let
    \begin{equation*}
        \cvar_{\alpha}(\mathsf{X}) \defeq \expect{~\mathsf{X} \mid \mathsf{X} \geq \text{the \((1-\alpha)\)-quantile value of the belief distribution}~},
    \end{equation*}

\begin{proposition}\label{prop:profit-cvar-lower-bound}
    Suppose that \( \priceR \in [\sqrt{g}, 1]\) and \( \priceL \in [\sqrt{1-g}, 1] \). Then,
    \begin{equation*}
        u(\priceR, \priceL) \geq \cvar_{\alpha}(\pt) + \cvar_{\beta}(\qt) - (\priceR + \priceL),
    \end{equation*}
    where
    \begin{align*}
        \alpha = \left( \frac{1-g}{1-\priceR} - \frac{g}{\priceR} \right)^{-1}; \quad
        \beta = \left( \frac{g}{1-\priceL} - \frac{1-g}{\priceL} \right)^{-1}.
    \end{align*}
\end{proposition}

\begin{proof}
We will use the variational form of \( \cvar \) depicted in the following theorem.

\begin{theorem}[{\citet[Theorem 1]{rockafellar2000optimization}}]
    \label{thm:variational}
    The conditional value-at-risk of a random variable has the following variational form:
    \begin{equation}
        \label{eq:cvar_variational}
        \cvar_{\alpha}(\mathsf{Z}) = \inf_{\rho \in \mathbb{R}} \left\{ \rho + \frac{\expect{(\mathsf{Z} - \rho)_{+}}}{\alpha} \right\}.
    \end{equation}
\end{theorem}

By using the variational form \cref{eq:cvar_variational}, we have
    \begin{align*}
        \expect{(\pt - \priceR)_+}
        &= \alpha \cdot \left( \priceR + \frac{\expect{(\pt - \priceR)_+}}{\alpha} - \priceR \right) \\
        &\geq \alpha \cdot \left( \inf_{\rho} \left\{ \rho + \frac{\expect{(\pt - \rho)_+}}{\alpha} \right\} - \priceR \right) \\
        &= \alpha \cdot \left( \cvar_{\alpha}(\pt) - \priceR \right).
    \end{align*}
    Analogously, $\expect{(\qt - \priceL)_+} \ge \beta \cdot \left( \cvar_{\beta}(\qt) - \priceL \right)$.

    One can verify that given the conditions of \( \priceR, \priceL\), we have that the terms satisfy:
    \begin{align*}
        \left( \frac{1-g}{1-\priceR} - \frac{g}{\priceR} \right)^{-1},
        \left( \frac{g}{1-\priceL} - \frac{1-g}{\priceL} \right)^{-1} \in [0, 1]
    \end{align*}
    As such, taking \( \alpha \) and \( \beta \) per the theorem, we have,
    \begin{align*}
        u(\priceR, \priceL)
        &=
        \left(\frac{1-g}{1-\priceR} - \frac{g}{\priceR}\right)\expect{(\pt-\priceR)_+} + \left(\frac{g}{1-\priceL} - \frac{1-g}{\priceL}\right)\expect{(\qt-\priceL)_{+}} \\
        &= \alpha^{-1}\expect{(\pt-\priceR)_{+}} + \beta^{-1}\expect{(\qt-\priceL)_{+}} \\
        &\geq \cvar_{\alpha}(\pt) + \cvar_{\beta}(\qt) - (\priceR + \priceL).\qedhere
    \end{align*}
\end{proof}

%% file: appendix/imprecise-belief.tex
\section{Lemma on Imprecise Bookmaker Belief}
\setcurrentname{Lemma on Imprecise Bookmaker Belief}
\label{app:imprecise_belief}

In the following, we present a formal Lemma to clarify our statement about a bookmaker having a belief of \eventR in the form of an interval $(g_-, g_+)$.
This imprecise belief allows the bookmaker to be less committed to their position about \eventR.

\begin{lemma}
    Suppose $g_-$ and $g_+$ are the bookmaker's lower and upper bound estimates of the ground truth probability $\prob{\eventR}$ such that $0\leq g_- \leq \prob{\eventR} \leq g_+ \leq 1$. Then, there exists $(\priceR, \priceL)$ such that the bookmaker's estimated profit is non-negative at time $t$, for any preference distribution.
\end{lemma}
\begin{proof}
    The proof follows similarly to the derivation of the bookmaker's expected profit \cref{eq:profit-cont}, as proven in \Cref{app: derivation of profit}.
    Consider the expected bookmaker's profit over the ground truth probability at time $t$:
    \[
     \ut^{\text{true}}(\priceR, \priceL) = \left(\frac{1-\prob{\eventR}}{1-\priceR} - \frac{\prob{\eventR}}{\priceR}\right)\expect{(\pt -\priceR)_{+}} 
        + \left(\frac{\prob{\eventR}}{1-\priceL} - \frac{1-\prob{\eventR}}{\priceL}\right)\expect{(\qt -\priceL)_{+}}.
    \]
     Since, $g_- \leq \prob{\eventR} \leq g_+$, the bookmaker could obtain the following lower bound:
    \begin{equation}
    \ut^{\text{true}}(\priceR, \priceL) \geq \left(\frac{1-g_+}{1-\priceR} - \frac{g_+}{\priceR}\right)\expect{(\pt -\priceR)_{+}} 
        + \left(\frac{g_-}{1-\priceL} - \frac{1-g_-}{\priceL}\right)\expect{(\qt -\priceL)_{+}} . \label{app-eq: lower-bound-imprecise}
    \end{equation}
    By setting $a \geq g_+$ and $1 -b \leq g_-$, we can conclude that both terms are nonnegative.
\end{proof}

We remark that \Cref{alg:SA} could still be applied to the scenario that the bookmaker has imprecise estimates of the probability. The reason is that the SA algorithm could be regarded as two separate stochastic approximation processes on $\Upsilon^R$ and $\Upsilon^L$ (see \Cref{eq:SA-FOC}). Hence we could work on those problems with $g_+$ and $g_-$ (as parameters of the function $G$) separately to optimise the lower bound \Cref{app-eq: lower-bound-imprecise}.

%% file: appendix/Discussion-uniqueness.tex
\section{Uniqueness of Maximisers}
\setcurrentname{Uniqueness of Maximisers}
\label{app: uniqueness-of-equilibrium}
Critical points of of the profit function \Cref{eq:profit-cont} are determined by the following equation
\begin{equation}
    \expect{\pt~|~\pt\geq \priceR} - \priceR = \frac{\priceR(1 - \priceR)(\priceR -g)}{\priceR^2 - 2g\priceR + g}, \label{app-eq: uniqueness}
\end{equation}
where the LHS is known as the Mean Residual Life (MRL) of the bettors' belief distribution. It is known that the MRL is decreasing when the belief distribution is log-concave \cite{al2008some}. If the root of the above equation is unique in the open interval $(0,1)$, then the root must be the unique maximiser of the profit function \Cref{eq:profit-cont}.

\begin{figure}[H]
    \begin{minipage}{0.49\textwidth}
    \centering
    \includegraphics[width =  1.\textwidth]{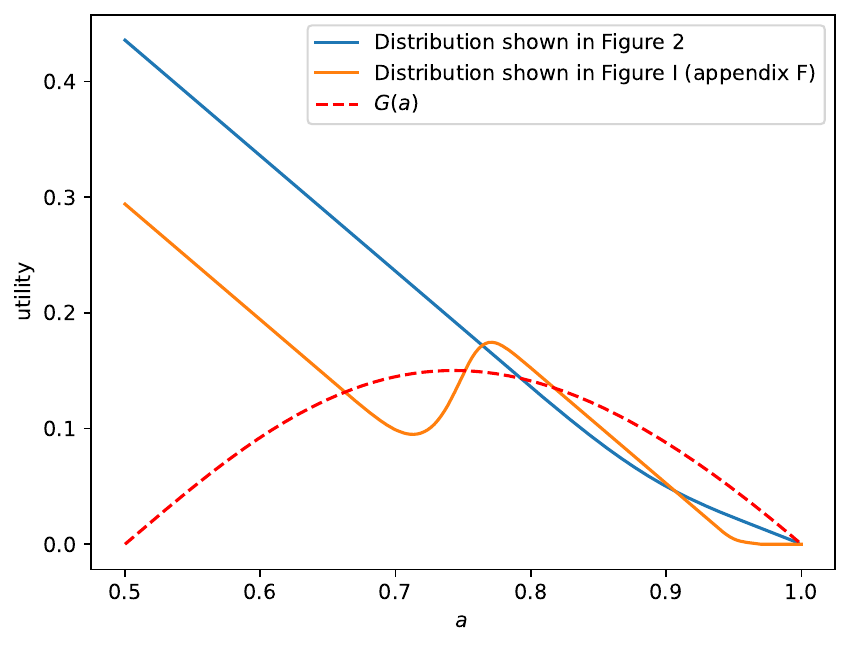}
    \end{minipage}
    \begin{minipage}{0.49\textwidth}
    \includegraphics[width =  1.\textwidth]{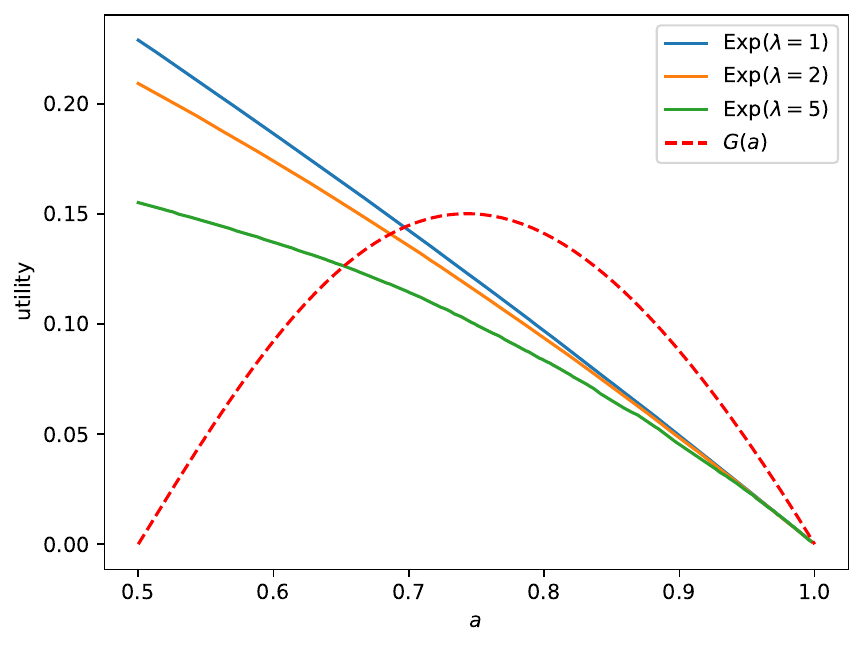}
    \end{minipage}
    \caption{Illustrations of the roots of \Cref{app-eq: uniqueness}. The dashed red line represents the value of RHS and all others represent the LHS with expectations taken \wrt different distributions. (left plot) Distributions used in \Cref{sec:Empirical} and \Cref{app: Empirical}. (right plot) Truncated exponential distributions with different parameters $\lambda \in \{1,2,5\}$.%
    }
    \label{app-fig:unique-1}
\end{figure}

We empirically examine the uniqueness of such roots for common distributions. \Cref{app-fig:unique-1} (Left) depicts the roots of the equation above. As demonstrated, the belief distribution used in the experiment of \Cref{sec:Empirical} corresponds to unique profit maximiser. Whereas the belief distribution corresponds to multiple maxmisers (\Cref{app: Empirical}) induces multiple roots of the above equation. Further, we plotted cases of (truncated) normal distribution and (truncated) exponential distributions. It turns out that all the cases we have tested admit unique profit maximisers.

\begin{figure}[H]
    \begin{minipage}{0.49\textwidth}
    \centering
    \includegraphics[width =  1.\textwidth]{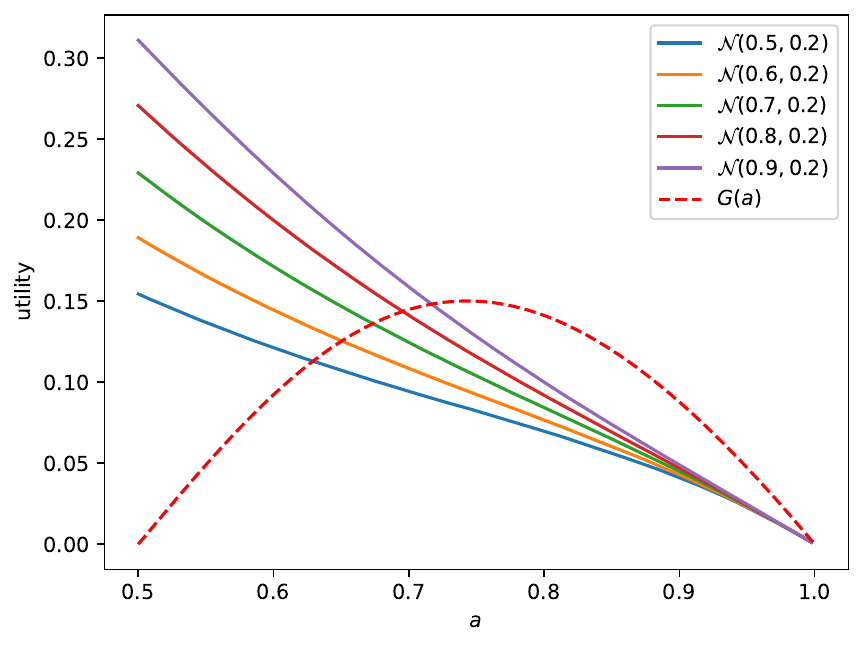}
    \end{minipage}
    \begin{minipage}{0.49\textwidth}
    \includegraphics[width =  1.\textwidth]{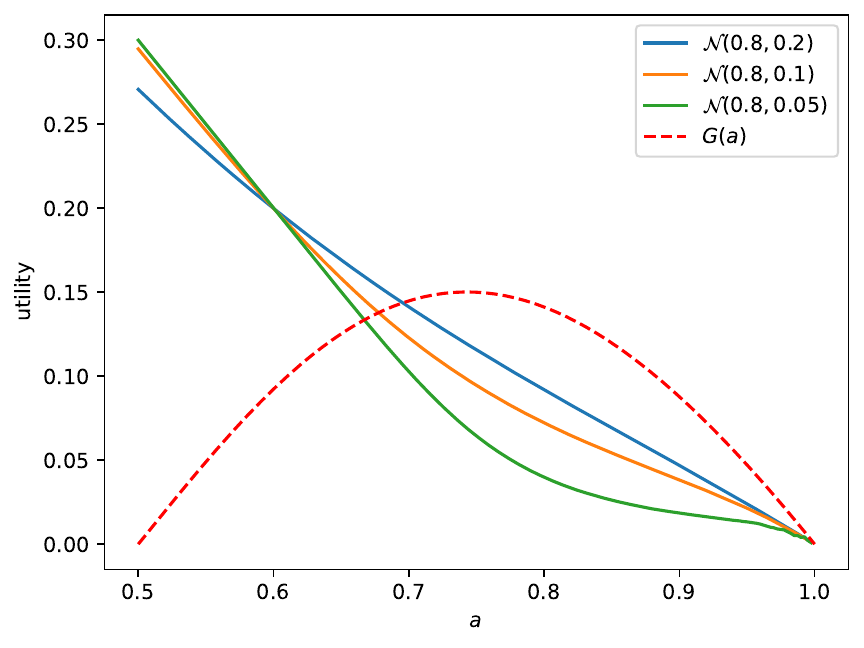}
    \end{minipage}
    \caption{Illustrations of the roots of \Cref{app-eq: uniqueness}. The dashed red line represents the value of RHS and all others represent the LHS with expectations taken \wrt different belief distributions. (left plot) Truncated Gaussian distributions with different means $\{0.5,0.6, 0.7, 0.8, 0.9\}$ and the same variance $0.2$. (right plot) Truncated Gaussian distributions with the same mean $0.2$ but different variances $\{0.2,0.1,0.05\}$.}
    \label{app-fig:unique-2}
\end{figure}

%% file: appendix/Incompatible.tex
\section{Incompatibility of Profit Maximisation and Prediction Aggregation}
\setcurrentname{Incompatibility of Profit Maximisation and Prediction Aggregation}
\label{app: incompatible-profit-prediction}

By \Cref{eq:profit-cont}, the expected profit when $g=0.5$ is
\[
\frac 12 \left(\frac{1}{1-\priceR} - \frac{1}{\priceR}\right)\cdot \expect{(\pt -\priceR)_{+}} + \frac 12 \left(\frac{1}{1-\priceL} - \frac{1}{\priceL}\right)\cdot \expect{(\qt -\priceL)_{+}}  ~.
\]
Let's focus on the first term. If $m-\Delta_1 \le \priceR \le m+\Delta_1$, the first term becomes
\[
\frac 12 \left(\frac{1}{1-\priceR} - \frac{1}{\priceR}\right)\cdot \expect{(\pt -\priceR)_{+}}
~=~ \frac 12 \left(\frac{1}{1-\priceR} - \frac{1}{\priceR}\right)\cdot \int_{\priceR}^{m+\Delta_1} (p-\priceR)\cdot \frac{1}{4\Delta_1}\,dp 
~=~ \frac{1}{16\Delta_1} \left(\frac{1}{1-\priceR} - \frac{1}{\priceR}\right) \left(\priceR - (m+\Delta_1)\right)^2~.
\]
If $0.5\le \priceR \le m-\Delta_1$, the first term becomes
\[
\frac 12 \left(\frac{1}{1-\priceR} - \frac{1}{\priceR}\right)\cdot \expect{(\pt -\priceR)_{+}}
~=~ \frac 12 \left(\frac{1}{1-\priceR} - \frac{1}{\priceR}\right)\cdot \int_{m-\Delta_1}^{m+\Delta_1} (p-\priceR)\cdot \frac{1}{4\Delta_1}\,dp \\
~=~ \frac 18 \left(\frac{1}{1-\priceR} - \frac{1}{\priceR}\right) (m-\priceR)~.
\]
We seek $\priceR = \priceR^\star$ that maximizes the expected profit over the interval $[0.5,m+\Delta_1]$.
There is no simple closed-form formula for $\priceR^\star$, so we numerically compute $\priceR^\star$ for different combinations of $m$ and $\Delta_1$.
For any fixed $m$, we report the possible range of $\priceR^\star$ in \cref{table:range-of-astar}.

%% file: appendix/additional-empirical.tex
\section{Additional Empirical Results}
\setcurrentname{Additional Empirical Results}
\label{app: Empirical}
\subsection{The Risk Balancing Algorithm} \label{app-alg: RB}
\input{figure/RiskBalance}
\subsection{SA under Multi-modal Distribution}
\begin{figure}[h]
    \centering
    \includegraphics[width = 1.\textwidth]{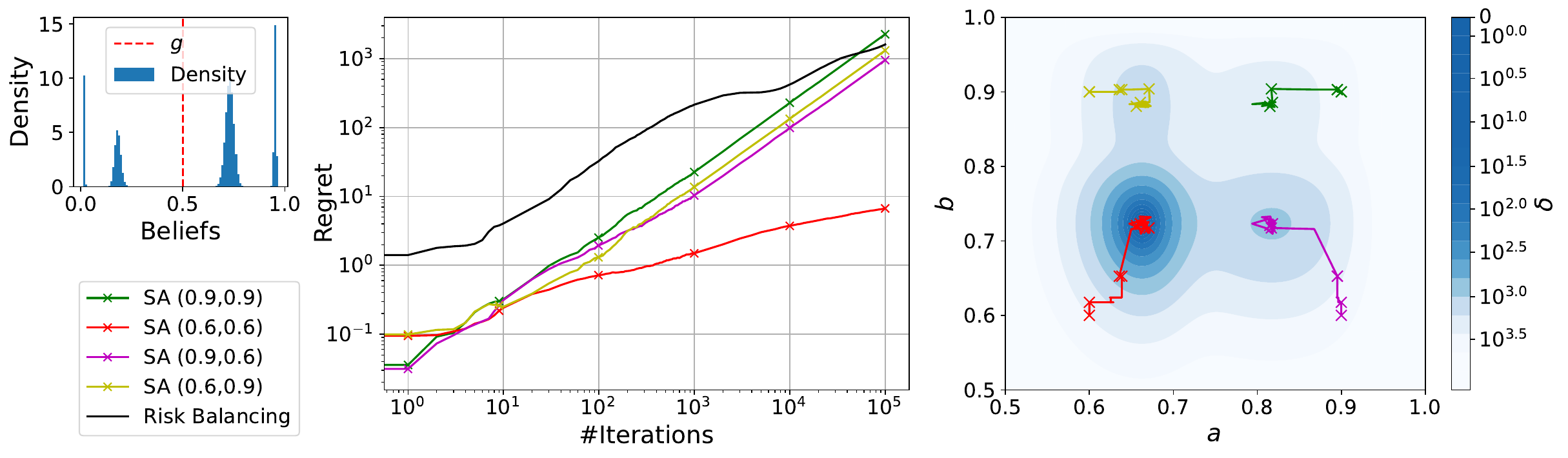}
    \caption{Simulation results of \Cref{alg:SA} algorithm with $100,000$ bettors. Left: The histogram distribution of bettors' beliefs. Middle: The regrets over $100,000$ iterations, data is collected every $1,000$ iterations. The points of \# iterations = $\{10^1, \ldots,10^5\}$ are marked with "$\times$". Right: Contour plot of $\delta(a,b) \defeq  u_{1:T}(\priceR^\star,\priceL^\star) - u_{1:T}(\priceR,\priceL)$, the darkness is of colour is associated with the value $\delta$ according to the color bar to the right.}
    \label{fig: regret-four-gau}
\end{figure}
 \Cref{fig: regret-four-gau} shows the cases when the maximiser of $u$ is not unique. From the right plot, we can see that these four processes, under different initialisations, will converge to four different maximisers respectively. From the middle plot, we could find that the regrets of processes converging to extremely ``bad'' maximisers (\ie green and purple) are increasing drastically and comparable to the risk balancing scheme. The process converges to 
 the global maximiser (red) and has regret $\leq 10^1$ throughout. 
Therefore, we conclude that under the regimes that the maximisers are not unique, proper initialisation is needed to attain the desired regret bound.

\subsection{Different Regret Definitions}

Other than the ``stochastic regret'' definition we used throughout the main text, another commonly used definition is the ``adversarial regret'' \cite{hazan2014beyond}, which could be defined as:%
\[
\begin{aligned}
    \textsc{Regret}^\text{adv}(T) &= \max_{(\priceR,\priceL): \priceR + \priceL \geq 1} \left\{\sum_{t=1}^T  \left(\frac{1-g}{1-\priceR} - \frac{g}{\priceR}\right)(\pt - \priceR)_+w_t+ \left(\frac{g}{1-\priceL} - \frac{1-g}{\priceL}\right)(\qt - \priceL)_+w_t\right\}  \\
    &~~~~~~~~~~~~~~~~~~~~~~~~- \sum_{t=1}^T  \left(\frac{1-g}{1-\priceR_t} - \frac{g}{\priceR_t}\right)(\pt - \priceR_t)_+w_t
        + \left(\frac{g}{1-\priceL_t} - \frac{1-g}{\priceL_t}\right)(\qt - \priceL_t)_+w_t,
\end{aligned}
\]
The adversarial regret is based on true wealth and beliefs of the bettors. We note that this quantity is not accessible for bookmakers, as the exact quantities of $w_t$ and $\pt$ are not assumed to be known by the bookmaker. However, we could still test our algorithm under such a benchmark to examine the efficiency.

\begin{figure}[h]
    \centering
    \includegraphics[width = 0.45\textwidth]{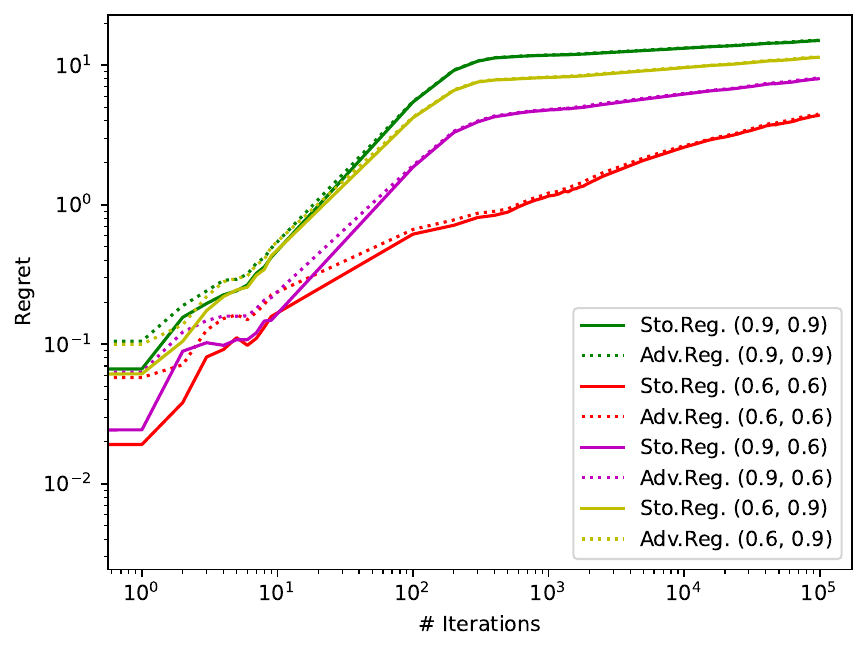}
    \caption{Stochastic and Adversarial Regrets under the belief distribution discussed in \Cref{sec:Empirical}}
    \label{fig:enter-label}
\end{figure}
We could identify that stochastic regret and adversarial regret are nearly indistinguishable after $10^3$ iterations, which justifies that our algorithm is also efficient in terms of adversarial regret.

%% file: figure/RiskBalance.tex
\begin{algorithm}[htb]
\caption{Risk Balancing Algorithm}\label{alg: RB}
\begin{algorithmic}[1]
\REQUIRE Initial price $\priceR_0, \priceL_0$, Learning rate $(\eta_{t+1})_{t\in \nn}$.
\STATE Initialise the total bets as $B_L = B_R = 0$
\FOR{each bettor entering the market at time $t$}
   \STATE Receive the bet placed by the bettor $\bet_t$.
   \IF{the bet is placed on \eventR}
      \STATE Update the total amount bet on \eventR.
       \begin{equation}
           B_R = B_R + \bet_t
       \end{equation}
   \ELSE
       \STATE Update the total amount bet on \eventL.
       \begin{equation}
           B_L = B_L + \bet_t
       \end{equation}
   \ENDIF
   \STATE  Update the prices following
   \[
   \priceR = \priceR + \eta_{t+1}(B_R - B_L),~~ \priceL = \priceL + \eta_{t+1}(B_L - B_R).
   \]
\ENDFOR
\end{algorithmic}
\end{algorithm}

%% file: appendix/Proof-for-section-2.tex
\section{Derivation of Kelly Bettor Strategy, \cref{eq:kelly-betting-rule}}
\setcurrentname{Derivation of Kelly Bettor Strategy, \cref{eq:kelly-betting-rule}}
\label{appendix_kelly}

Consider the problem 
\[
\max_{v\geq 0} \varphi_{t}^{\priceR}(v) = \max_{v\geq 0} \left\{\pt \log\left(\wt + \frac{1-\priceR}{\priceR} v \right) + \qt\log(\wt - v )\right\}.
\]
Let $v_\priceR^\star = \argmax_{v\geq 0} \varphi_{t}^{\priceR}(v)$. It is straightforward to verify that this problem is concave. Hence, by equalising the gradient to $0$, we have
\[
\frac{\partial \varphi_{t}^{\priceR}(v_\priceR^\star)}{\partial v} = \pt\cdot\frac{\frac{1-\priceR}{\priceR}}{\wt + \frac{1-\priceR}{\priceR} v_\priceR^\star} - \qt\cdot\frac{1}{\wt - v_\priceR^\star } = 0.
\]
This implies \(v_\priceR^\star = \frac{\pt - \priceR}{1 - \priceR}w_t\) if $\pt > a$ and $v_\priceR^\star =0$ otherwise. By symmetry, $v_\priceL^\star \defeq \argmax_{v\geq 0} \varphi_{t}^{\priceL}(v) =  \frac{\qt - \priceL}{1 - \priceL}w_t$ if $\qt > b$ and $v_\priceL^\star = 0$ otherwise. This implies the result. 

\begin{figure}[H]
    \begin{minipage}{0.4\textwidth}
    \centering
    \includegraphics[width =  1.\textwidth]{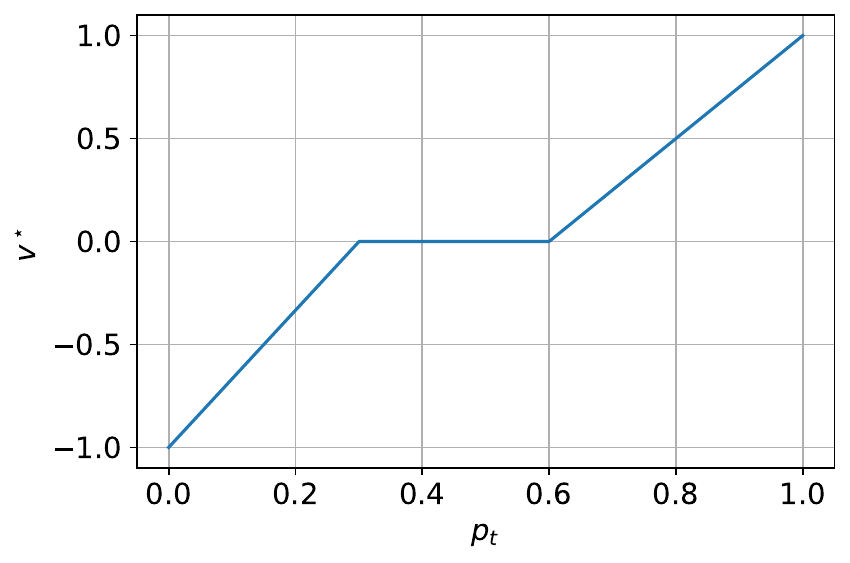}
    \end{minipage}
    \begin{minipage}{0.6\textwidth}
    \includegraphics[width =  1.\textwidth]{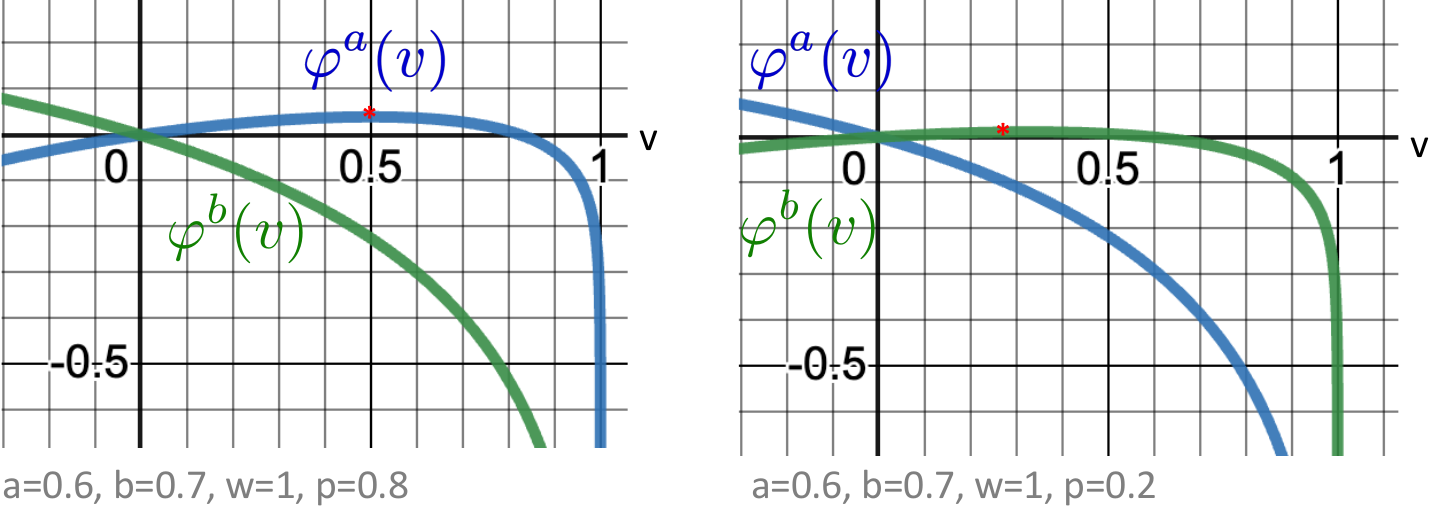}
    \end{minipage}
    \caption{An illustration of Kelly Bettor strategy with prices $\priceR = 0.6$ and $\priceL = 0.7$, bettor wealth $\wt=1$. (left plot) Optimal investment strategy $\vt^*$ is a function of bettor belief $\pbettor=\pt$. (middle and right plots) An illustration of the expected log wealth function for betting on either side  $\varphi^\priceR(v)$,  $\varphi^\priceL(v)$, with their maximum marked in red. At $\pt=0.8$, the optimal betting amount is $v^*=0.5$ for event \eventR; at $\pt=0.2$, the optimal betting amount is $v^*=0.33$ for event \eventL. }
    \label{app-fig:Kelly-strategy}
\end{figure}

One example of the Kelly Bettor strategy is illustrated in \cref{app-fig:Kelly-strategy}, where positive values are the amounts bet on \eventR, and negative values are the amounts bet on \eventL.

\section{Derivation of the Bookmaker's Expected Profit, \cref{eq:profit-cont}}
\setcurrentname{Derivation of the Bookmaker's Expected Profit, \cref{eq:profit-cont}}
\label{app: derivation of profit}

The bookmaker's {expected profit at time $t$} is equal to the {expected} amount of payin minus the {expected} amount of payout. Since the wealth of bettors is independent of the beliefs, we first assume the bettor's wealth is $1$. 

If event \eventR happens, by \cref{eq:kelly-betting-rule}, the amount of payin minus payout is
\[
\begin{aligned}
&~~~\int_{\priceR}^1 \frac{p - \priceR}{1 - \priceR}f(p)\ dp + \int_{0}^{1-\priceL} \frac{1- p - \priceL}{1 - \priceL}f(p)\ dp - \frac{1}{\priceR}\int_{\priceR}^1 \frac{p - \priceR}{1 - \priceR}f(p)\ dp \\
& = \frac{1}{\priceR} \int_{\priceR}^1 (\priceR - p)f(p)\ dp + \int_{0}^{1-\priceL} \frac{1- p - \priceL}{1 - \priceL}f(p) \ dp.
\end{aligned}
\]

Similarly, if the event \eventL happens, the amount of payin minus payout is
\[
\begin{aligned}
&~~~\int_{\priceR}^1 \frac{p - \priceR}{1 - \priceR}f(p)\ dp + \int_{0}^{1-\priceL} \frac{1- p - \priceL}{1 - \priceL}f(p)\ dp  - \frac{1}{\priceL} \int_{0}^{1-\priceL} \frac{1- p - \priceL}{1 - \priceL}f(p)\ dp \\
& = \int_{\priceR}^1 \frac{p - \priceR}{1 - \priceR}f(p)\ dp + \frac{1}{\priceL}\int_{0}^{1-\priceL} (\priceL + p -1)f(p) \ dp.
\end{aligned}
\]
We could get the result directly by combining the terms and noticing the facts
\[
\begin{aligned}
-\expect{(\pt- \priceR)_+} &=  \int_{\priceR}^1 (\priceR - p)f(p)\ dp, \\
-\expect{(1 - \pt -\priceL)_+} &= \int_{0}^{1-\priceL} (\priceL + p -1)f(p) \ dp.
\end{aligned}
\]

\section{Proofs for \Cref{sec:equilibria}}
\setcurrentname{Missing Proofs for \Cref{sec:equilibria}}
\label{app: missing-proof-sec3}

\subsection{Proof of \Cref{lem:existance-of-maximiser}}
 \setcurrentname{Proof of \Cref{lem:existance-of-maximiser}}
\label{app: proof-of-existance-of-equilibria}
\existanceOfMaximiser*
\begin{proof}
    
    Since $T$ is fixed, it suffices to consider the static case where $u \defeq u_t$. To see $u$ is upper-bounded, for the first term we have 
    \[
    \left(\frac{1-g}{1-\priceR} - \frac{g}{\priceR}\right)\expect{(\pt-\priceR)_+}
    = \frac{\priceR - g}{(1 - \priceR)\priceR}\int_{\priceR}^1 (p - \priceR)f(p) \ \text{d}p
    \leq 1 - \frac{g}{\priceR},
    \]
    where the last inequality follows from $p\leq1$. Similarly, 
    \[
    \left(\frac{g}{1-\priceL} - \frac{1-g}{\priceL}\right)\expect{(1-\priceL-\pt)_+}
    = \frac{\priceL - (1-g)}{(1-\priceL)\priceL}\int_0^{1-\priceL} (1-\priceL-p)f(p) \ \text{d}p
    \leq 1 .
    \]
    Therefore, $u$ is upper-bounded. Next, to show the maximiser exists, we first recall that the domain of the profit function is the set \(D=\)\{\(0 < \priceR < 1\), \(0 < \priceL < 1\), \(\priceR + \priceL \geq 1\)\}. Define a sequence $(x_k) \subset D$ such that $u(x_k) \rightarrow \sup_{x\in D} u(x)$. Since the sequence is bounded, there exists $x^*$ which is the limit of a convergent subsequence. We only need to show $x^*\in D$. Indeed, $x^*$ is in the closure of $D$. Denote the closure of $D$ as $\bar{D}$, we notice that $\bar{D} \setminus D = \{(a,b): a=1 \text{ or }b=1, a+b \geq 1\}$. We can see that the limit
    \[
    \lim_{a\rightarrow 1} \frac{(1-g)\expect{(\pt-\priceR)_+}}{1-a} = \lim_{a\rightarrow 1} \frac{(1-g) \cdot \int_{a}^1 1 - F(x)~\text{d}x}{1-a}
     = \lim_{a\rightarrow 1} \frac{(1-g) \cdot (F(a) -1)}{-1} = 0.
    \]
    Therefore,  
     \[
    \lim_{\priceR \rightarrow 1}\left(\frac{1-g}{1-\priceR} - \frac{g}{\priceR}\right)\expect{(\pt-\priceR)_+}
    = \lim_{\priceR \rightarrow 1}- \frac{g}{\priceR}\expect{(\pt-\priceR)_+} =0.
    \]
    However, it is clear that the term $ \left(\frac{1-g}{1-\priceR} - \frac{g}{\priceR}\right)\expect{(\pt-\priceR)_+}$ could achieve some positive value when $g<a<1$. Thus, we can rule out the cases that any sequences $(x_k)_{k\in \nn} = (a_k, b_k)_{k\in \nn}$ such that $a_k \rightarrow 1$ will converge to some maximiser. Exactly the same arguments could also rule out the cases that sequences with $b_k \rightarrow 1$ will converge to some maximiser, which concludes that the maximiser should exist in the set $D$.
\end{proof}

\subsection{Proof of \Cref{lem:char-maximiser}}
\setcurrentname{Proof of \Cref{lem:char-maximiser}}
\label{app: proof-of-char-maximiser}
\charMaximiser*
\begin{proof}
For notational simplicity, we denote the first term of $u$ as $\Pi_1$ and the second term of $u$ as $\Pi_2$:
\[
\begin{aligned}
    \Pi_1 &=
        \left( \frac{1-g}{1-\priceR} - \frac{g}{\priceR}\right)\expect{(\pt - \priceR)_+} \\
    \Pi_2 &= 
        \left( \frac{g}{1-\priceL}-\frac{1-g}{\priceL} \right)\expect{(1 - \priceR - \pt)_+}.
\end{aligned}
\]
We will prove by contradiction. Since $\priceR + \priceL \geq 1$, the other possibilities of the ordering could be $1-\priceL^* \leq \priceR^* \leq g$ or $g\leq 1 - \priceL^* \leq \priceR^*$. Suppose $1-\priceL^* \leq \priceR^* \leq g$, then we must have $\Pi_1 \leq 0$. However, if we increase $\priceR^*$ to $\widetilde{\priceR^*}$ such that $\widetilde{\priceR^*}> g$. We will have $$\Pi_1(\widetilde{\priceR^*})> 0 \geq \Pi_1(\priceR^*).$$Therefore, $(\widetilde{\priceR^*}, \priceL^*)$ is still a maximiser. Hence, this leads to a contradiction since $(\priceR^*, \priceL^*)$ cannot be the maximiser. With the same arguments, we can also rule out the other case.
\end{proof}

%% file: appendix/Profit-Margin-Proofs.tex
\subsection{Proof of \Cref{prop:profit-margin-preference-deviation}}
\setcurrentname{Proof of \Cref{prop:profit-margin-preference-deviation}}
\label{app: Profit-margin-squared-difference}
\profitMargin*
\begin{proof}

We consider the case that $\priceR + \priceL = 1$ since the lower bound is still valid when we ease the assumption to $\priceR + \priceL \geq 1$. Consider the price $\priceR = \frac{\expect{\pt} + g}{2}$, the profit becomes
\[
u(\priceR, 1-\priceR) = \frac{g - \expect{\pt}}{2} \cdot \frac{g  - \expect{\pt}}{2} \cdot \frac{1}{(1-\priceR)\priceR}
\]
where the term $\frac{1}{(1-\priceR)\priceR}$ is minimised when $\priceR = \frac{1}{2}$. Hence, we could conclude the result.

\end{proof}

\subsection{Proof of \Cref{lem:Stochastic-Dominance-Profit-Margin}}
\setcurrentname{Proof of \Cref{lem:Stochastic-Dominance-Profit-Margin}}
\label{app: Proof-Stochastic-Dominance-Profit-Margin}
\StochasticDominance*
\begin{proof}

Since $F_1$ SOSD over $F_2$, we have
\[
 \int_{0}^x F_1(w)\ dw <  \int_{0}^x F_2(w)\ dw.
\]
Substracting both sides with $\int_{0}^1 F_1(w)\ dw = \int_{0}^1 F_2(w)\ dw$ we have
\[
\int_{x}^1 F_1(w)\ dw >  \int_{x}^1 F_2(w)\ dw.
\]
For the profit function, we note that
\[
\expect{(\pt-\priceR)_+} = \int_{\priceR}^1 (p - \priceR)f(p)\ dp  = 1 - \priceR -\int_{\priceR}^1 F(p)\ dp,
\]
where $F$ is the cdf. Hence,
\[
\mathbb{E}_1[(\pt-\priceR)_+] = 1 - \priceR -\int_{\priceR}^1 F_1(p)\ dp < 1 - \priceR -\int_{\priceR}^1 F(p)\ dp = \mathbb{E}_2[(\pt-\priceR)_+].
\]
Similarly, for the second term, we have
\[
\mathbb{E}_1[(1-\priceL-\pt)_+] = \int_{0}^{1-\priceL} F_1(p) \ dp < \int_{0}^{1-\priceL} F_2(p) \ dp  = \mathbb{E}_2[(1-\priceL-\pt)_+].
\]
Since $1-\priceL < g < \priceR$, the coefficients of expectations in the profit function are positive. Therefore, it is clear that $ u_1(\priceR, \priceL) < u_2(\priceR, \priceL)$.

For the maxima of profit, suppose $(\priceR^*, \priceL^*)$ is the maximiser of $u_1$. By \Cref{lem:char-maximiser}, we know that $1- \priceL^* < g < \priceR^*$. Therefore, 
$u_1^* = u_1(\priceR^*, \priceL^*) < u_2(\priceR^*, \priceL^*) \leq u_2^*$.

\end{proof}

%% file: appendix/online-algorithm-general.tex
\section{Proof of \Cref{lem:FOC-General Case}}
\setcurrentname{Proof of \Cref{lem:FOC-General Case}}
\label{app: FOC-General-case}
\FocGeneralCase*
\begin{proof}
By symmetry, we will focus on the right hand side price $\priceR$. We notice that 
\begin{equation}
\begin{aligned}
\expect{(\pt - \priceR)_+} &= \prob{\pt \geq \priceR}\cdot \expect{(\pt - \priceR)_+ ~|~ \pt \geq \priceR} + \prob{\pt \leq \priceR}\cdot \expect{(\pt - \priceR)_+ ~|~ \pt \leq \priceR}\\
&= \prob{\pt \geq \priceR}\cdot \expect{\pt - \priceR ~|~ \pt \geq \priceR} \\
&= \prob{\pt \geq \priceR}\cdot (\expect{\pt ~|~ \pt \geq \priceR} - \priceR) \\
&= (1-F(a))\cdot (\expect{\pt ~|~ \pt \geq \priceR} - \priceR).
\end{aligned}\label{appEq:decomposing expectation}
\end{equation}
On the other hand, $\expect{(\pt - \priceR)_+} = \int_{\priceR}^1 (\pt-\priceR) f(\pt)\,d\pt$, so by the Leibniz integration rule,
$\frac{\partial \expect{(\pt - \priceR)_+}}{\partial \priceR} = -\int_{\priceR}^1 f(\pt)\,d\pt = F(a)-1$. Thus,
the derivative of the profit function is
\[
        \frac{\partial u}{\partial \priceR} = \frac{(\prefCDF(\priceR) - 1)(\priceR -g)}{\priceR(1-\priceR)} + \frac{\expect{(\pt - \priceR)_+}(\priceR^2 - 2g\priceR +g)}{\priceR^2(1 - \priceR)^2}.
\]
Hence, for critical points, we have
\[
    \priceR(1 - \priceR)(\priceR -g)(\prefCDF(\priceR) -1) + \expect{(\pt - \priceR)_+}(\priceR^2 - 2g\priceR +g) = 0.
\]
Further, by \Cref{appEq:decomposing expectation}, we can reformulate above equation to
\[
    \expect{\pt~|~\pt\geq \priceR} - \priceR = \frac{\priceR(1 - \priceR)(\priceR -g)}{\priceR^2 - 2g\priceR + g}.
\]

\end{proof}

\section{Proof of \Cref{thm:SA-convergence}}
\setcurrentname{Proof of \Cref{thm:SA-convergence}}
\label{app: SA-convergence proof}
We first restate the theorem:
\saConvergence*
The proof of \Cref{thm:SA-convergence} has three main steps. First, we show that the update rule in \Cref{alg:SA} is indeed a stochastic approximation process as in \Cref{def:SA-process}. Second, we leverage the convergence results of stochastic approximation process from existing literature~\cite{Gadat2018} to show that \Cref{alg:SA} converges toward a critical point. Finally, we use two theorems of \citet{pemantle1990nonconvergence,renlund2010generalized} to show that the critical point is a local maximizer almost surely.
The proofs below focus on \Cref{eq:price_update_pos} where the bettors bet on the positive side (i.e. $\pt > \priceR_t$).
The other side is symmetric.

\subsection{Step 1: \Cref{alg:SA} is a Stochastic Approximation Process}

The update rule \Cref{eq:price_update_pos} can be written in the form of
\[
\priceR_{t+1} = \priceR_t - \eta_{t+1}(h(\priceR_t) + M_{t+1}),
\]
where $h(\priceR_t) =  \priceR_t + G(\priceR_t)-\expect{\pt~|~\pt\geq \priceR_t}$ and $M_{t+1} =  \expect{\pt~|~\pt\geq \priceR_t} - \widehat{p_t}$. 
We will show that this update rule is a stochastic approximation process as in \Cref{def:SA-process}.
First, we check that $|h(\priceR_t)|$ and $|M_{t+1}|$ are bounded for every $t\in \nn$. 

\begin{lemma}
    Assume the process $(\priceR_t)_{t\in \nn}$ is generated by \Cref{alg:SA} and stays in the region $[g,1]$. Then, 
    \[
    |h(\priceR_t)| \leq 2,~~~|M_{t+1}| \leq 1,
    \]
    for every $t\in\nn$. \label{app-lem: noise-bounded}
\end{lemma}
\begin{proof}
First of all,
\[
\left|h(\priceR_t)\right| = \left|\priceR_t + G(\priceR_t)-\expect{\pt~|~\pt\geq \priceR_t}\right| 
\leq  \left|G(\priceR_t)\right| + \left|\priceR_t -\expect{\pt~|~\pt\geq \priceR_t}\right|.
\]
For any $\priceR_t\in [g,1]$, we have
\[
G(\priceR_t) = \frac{\priceR_t(1-\priceR_t)(\priceR_t-g)}{\priceR_t^2 - 2g\priceR_t + g} = \frac{\priceR_t(1-\priceR_t)(\priceR_t-g)}{\priceR_t(\priceR_t-g) + g(1-\priceR_t)} \leq 1- \priceR_t \leq 1.
\]
Also, both $\priceR_t$ and $\expect{\pt~|~\pt\geq \priceR_t}$ are bounded within $[0,1]$, so their absolute difference is at most 1.
This concludes that $\left|h(\priceR_t)\right|\le 2$.

To bound $|M_{t+1}|$, we notice that by the assumption $\bet_t \leq \avgwealtht$, we have 
\[
0\leq \widehat{p_t} = \frac{(1 - \priceR_t)\bet_t}{\avgwealtht} + \priceR_t \leq 1 - \priceR_t + \priceR_t = 1.
\]
Hence, both $\widehat{p_t}$ and $\expect{\pt~|~\pt\geq \priceR_t}$ are bounded within $[0,1]$, so their absolute difference is at most $1$. 
\end{proof}

It remains to show that $\expect{M_{t+1}\mid \calF_t} = 0$, which is established in the next lemma.

\begin{lemma}
    Let $(\calF_t)_{t\in \nn}$ be the filtration that the SA process generated by \Cref{alg:SA} adapted to. For every $t\in \nn$, we have 
    \[
     \expect{\widehat{\pt} \mid \calF_t} = \expect{\pt \mid \pt \geq \priceR_{t}}.
    \]
    \label{lem:SA-well-defined}
\end{lemma}
\begin{proof}
We could notice that
\[
\expect{\widehat{p_t}\mid\calF_t} = \expect{\frac{(1 - \priceR_t)\bet_t}{\avgwealtht} + \priceR_t \mid \calF_t},
\]
where $\bet_t = \frac{p_t - \priceR_t}{1- \priceR_t}w_t$ from the Kelly bettor's rule. Hence, 
\[
\begin{aligned}
    \expect{\frac{(p_t - \priceR_t)w_t}{\avgwealtht} + \priceR_t \mid \calF_t} &= \expect{\priceR_t(1 - \frac{w_t}{\avgwealtht}) + p_t \mid \calF_t}\\
    &= \expect{p_t\mid\mathcal{F}_t} + \priceR_t\frac{\expect{\avgwealtht - w_t}}{\avgwealtht}\\
    &= \expect{\pt\mid \pt\geq \priceR_t} + \priceR_t\frac{\expect{\avgwealtht - w_t}}{\avgwealtht} \\
    &= \expect{\pt\mid \pt\geq \priceR_t}.
\end{aligned}
\]
where the third equality follows from the fact that, at time $t$, $p_t$ is sampled from the belief distribution conditioned on the event $\{p_t > \priceR_t\}$ and the last equlity holds as $\avgwealtht$ is the unbiased estimator of $w_t$.
\end{proof}
\subsection{Step 2: Convergence to Critical Points} \label{app:SA-learning-rate}
It is known that SA dynamics converge under some assumptions. In our case, we immediately identify that the profit function $\profit$ is a Lyapunov function of the dynamic. We will adopt the following technical result.
\begin{theorem}[{\citet[Corollary 2.3.2]{Gadat2018}}]
    Let $\Omega$ be a convex set in $\mathbb{R}$. Suppose $(X_t)_{t\in \nn} \subset \Omega$ is a stochastic approximation process defined in \Cref{def:SA-process}.
    Let $V$ be twice differentiable and $L$-smooth function such that for any $x\in \Omega$,
    \[
    \nabla V(x)\cdot h(x)\geq 0, ~~~ |h(x)|^2 + |\nabla V(x)|^2 \leq C(1+V(x))
    \]
    for some positive constant $C$;  furthermore, the noise terms satisfies
    \[
    \expect{|M_t|^2 | \calF_{t-1}} \leq C(1 + V(X_{n-1})) ~~~ \forall t\in \nn.
    \]
    Assume that the set $\{x \mid V(x) = v\} \cap \{ x \mid \nabla V(x)\cdot h(x) = 0\}$ is finite for every $v \in \mathbb{R}$. Then, $X_t$ converges towards $X_{\infty}$ almost surely and $\nabla V(X_\infty) \cdot h(X_\infty) = 0$.
\end{theorem}
Therefore, by restricting the learning rate $\eta_t$ so that $a_t$ remains bounded within the interval $[g,1]$, we can use the above theorem to establish the convergence result. 
We prove the following lemma to show that the process is restricted in the region $\priceR_t \in (g,1)$ and $\priceL \in (1-g,1)$ under the assumptions of \Cref{thm:SA-convergence}.
\begin{lemma}
    The process $(\priceR_t)_{t\in \nn} \subset  (g,1)$ for all $t\in \nn$ if and only if $a_0 \in [g,1)$ and
    \[
      \eta_t < \min\left\{1, \frac{1}{L_G}\right\},
    \]
    where $L_G$ is the Lipschitz constant of the function $G$ defined in \Cref{lem:FOC-General Case}. 
\end{lemma}

\begin{proof}
We will prove by induction. For the base case, $a_0$ is picked within $[g, 1)$ on purpose. Next, we assume $\priceR_t \in [g,1)$.
We note that
\[
\begin{aligned}
     \priceR_{t+1} &=  \priceR_t + \eta_t(\widehat{p_{t}} - \priceR_t - G(\priceR_t)) \\
     & =  \priceR_t + \eta_t\left(\frac{(p_t - \priceR_t)w_t}{\avgwealtht} - G(\priceR_t)\right)\\
     & \geq \priceR_t - \eta_t\cdot G(\priceR_t).
\end{aligned}
\]
where the inequality follows from the fact that $p_t \geq \priceR_t$. If $\eta_t \leq \frac{1}{L_G}$, then 
\[
\eta_t \leq \frac{1}{L_G} \leq \frac{\priceR_t - g}{G(\priceR_t) - G(g)} = \frac{\priceR_t - g}{G(\priceR_t)}.
\]
It follows that 
\[
\priceR_{t+1}\geq \priceR_t - \eta_t\cdot G(\priceR_t) \geq g.
\]

Since $\widehat{p_t}\leq 1$, by the fact that $\eta_t \leq 1$, we can conclude that
\[
\begin{aligned}
    \priceR_{t+1} &=  \priceR_t + \eta_t(\widehat{p_{t}} - \priceR_t - G(\priceR_t))\\
    &\leq \priceR_t + \eta_t(1 - \priceR_t - G(\priceR_t))\\
    & \leq 1 -G(\priceR_t) \leq 1.\qedhere
\end{aligned}
\]
\end{proof}

\subsection{Step 3: Non-convergence to local minimizers or boundary points} \label{app:SA-non-convergence}
 The last part to conclude the proof is to classify the critical points and show that the SA dynamic will avoid local minimisers with probability $1$. First, we utilise the following theorem to exclude the probability that the dynamic will converge to interior local minimisers.
 \begin{theorem}[{\citet[Theorem 1]{pemantle1990nonconvergence}}]
     Let $(X_t)_{t\in \nn}$ be a stochastic approximation process defined in \Cref{def:SA-process}. Suppose $(X_t)_{t\in \nn} \subset \text{int~}\Omega$ for some set $\Omega$, and $h(p) = 0$ for some $p \in \Omega$. Let $\calN_p$ be a neighbourhood of $p$. Assume that, for some constants $c_1, c_2, c_3,c_4 >0$, the following conditions are satisfied whenever $X_t \in \calN_p$ and $t$ is sufficiently large:
     \begin{itemize}
         \item $h$ is twice differentiable;
t         \item $h(x)(x - p) < 0$ for any $x\neq p$ and $x$ is close to $p$;
         \item $\frac{c_1}{t} \leq \eta_t \leq \frac{c_2}{t}$;
         \item $\expect{|M_{t}|} \geq c_3$; and
         \item $|M_t| \leq c_4$ almost surely.
     \end{itemize}
     Then $\prob{X_{t} \rightarrow p} = 0$. \label{thm:SA-non-convergence-interior}
 \end{theorem}

Since the last assumption of \Cref{thm:SA-convergence} rules out the possibility of saddle points, %
the critical point is either a local maximiser or a local minimiser. %
Applying \Cref{thm:SA-non-convergence-interior} will rule out the possibility of convergence to local minimisers. Next, we are still left to check that the dynamic will not converge to boundaries, where we note that the boundary points $\{0,1\}$ are indeed critical points of the profit function, which could not be handled directly by \Cref{thm:SA-non-convergence-interior}. To demonstrate non-convergence to such boundaries, we will rely on Renlund's \cite{renlund2010generalized} result and control the convergence to the boundary by inspecting the intrinsic characteristics of the dynamics.

We will focus on the non-convergence to $1$. It is immediate that $1$ is a critical point. Moreover, the profit is $0$ when $\priceR = 1$. We formally state the main technical tool below.

\begin{theorem}[{\citet[Theorem 3]{renlund2010generalized}}]
 Suppose that $(X_t)_{t\in \nn} \subset (a,b)$ is a SA process defined in \Cref{def:SA-process}. Assume that for any $p\in \{a,b\}$, $h(p) = 0$ and %
 $h(x)(x - p) < 0$ for any $x\neq p$ and $x$ is close to $p$. Also, assume there exists positive constants $K_1, K_2$ such that
 \[
  \expect{M_{t+1}^2 | \calF_t} \leq K_1|X_t - p|,
 \]
 \[
  [h(x)]^2 \leq K_2|x - p|,
 \]
 \begin{equation}
     t\cdot|X_t - p| \rightarrow \infty,~~\text{as}~t\rightarrow\infty. \label{app-eq:non-convergence-condition}
 \end{equation}
 Then $\prob{X_t\rightarrow p} = 0$.
\end{theorem}
We start by checking $h(x)(x - 1) <0$ when $x$ is near $1$. It is equivalent to showing $h(x) > 0$ when $x$ is near $1$. Recall that $h(x) = x + G(x) - \expect{\pt\mid \pt\geq x} = \Upsilon^R(x)$, which is a continuous function since the \pdf $f$ is differentiable. By the third assumption of \Cref{thm:SA-convergence}, $\Upsilon^R(x)$ has finitely many roots in the interval $(0,1)$. Thus, for all $x$ between the largest root and $1$, $\Upsilon^R(x)$ must have the same sign. By our calculations in \Cref{app: FOC-General-case}, $\Upsilon^R(x)$ has opposite sign of $\frac{\partial u(x)}{\partial x}$. Also, observe that in \cref{eq:profit-cont}, when $x$ is near 1, $\left(\frac{1-g}{1-x} - \frac{g}{x}\right)\expect{(\pt -x)_{+}}$ is strictly positive while its value is zero when $x\nearrow 1$. Combining all of the above, $\frac{\partial u(x)}{\partial x}$ must be negative when $x$ is near $1$, implying $h(x) = \Upsilon^R(x)$ must be positive.

Then we will mainly focus on verifying \Cref{app-eq:non-convergence-condition} since the other assumptions in the above theorem are immediately satisfied. To start, we have the following lemma. %
\begin{lemma}\label{lem:quadratic-upper-bound}
For $x\in [g,1]$,
\[
1-x - G(x) \leq C \cdot (1-x)^2
\]
for some constant $C$.
\end{lemma}
\begin{proof}
    By direct calculation, we could see that it is equivalent to show that
    \[
    \frac{g}{x^2 - 2gx + g} \leq C.
    \]
    For $C\geq \frac{1}{1-g}$, we have $(1- \frac{1}{C})g \geq g^2$. Hence, $x^2 - 2gx + (1-\frac{1}{C})g \geq  x^2 - 2gx + g^2 \geq 0$. It concludes the proof.
\end{proof}
Next, we are ready to verify that \eqref{app-eq:non-convergence-condition} is satisfied for the dynamic generated by \Cref{alg:SA}.
\begin{lemma}
    Let $(\priceR_t)_{t\in \nn}$ be the process generated by \Cref{alg:SA}, we have
    \[
     t\cdot|\priceR_t - 1| \rightarrow \infty,~~\text{as}~t\rightarrow\infty.
    \]
\end{lemma}
\begin{proof}
  First, since $\bet_t\leq \avgwealtht$, by \Cref{lem:quadratic-upper-bound} we have
    \[
    \begin{aligned}
        \priceR_{t+1} &=  \priceR_t + \eta_t(\widehat{p_{t}} - \priceR_t - G(\priceR_t))\\
        &\leq \priceR_t + \eta_t(1 - \priceR_t - G(\priceR_t))\\
        & \leq \priceR_t + C\eta_t(1-\priceR_t)^2.
    \end{aligned}
    \]
    For some large enough $T$, we have $\frac{ C\gamma }{t+m}\leq 1$ for all $t\ge T$. Define a sequence $(x_t)_{t\in \nn}$ such that $x_1 = \priceR_T$ and $x_{t+1} = x_t + \frac{1}{t+T+m+1}(1-x_t)^2$. It follows that $1- \priceR_{t+T} \geq 1 - x_t$ for any $t\geq 1$. %
    Hence, it suffices to show that $t\cdot(1-x_t) \rightarrow \infty$. Let $y_t = 1 - x_t$, We hypothesise that $y_t \geq \frac{1}{\sqrt{t}}y_1$. We will prove inductively. The base case is trivial. For the step case, we assume that $y_t \geq \frac{1}{\sqrt{t}}y_1$, we aim to show that $y_{t+1} \geq \frac{1}{\sqrt{t+1}}y_1$. Notice that,
    \[
    (t+1)\cdot t^{1/2} - (t+1)^{1/2}\cdot t \geq y_1,
    \]
    for $y_1$ sufficiently small. And the LHS is increasing as $t$ increases. Hence,
    \[
    \frac{1}{t+1}\cdot \frac{1}{t}\cdot y_1^2 \leq \left(\frac{1}{t^{1/2}} - \frac{1}{(t+1)^{1/2}}\right)\cdot y_1.
    \]
    Therefore, by inductive hypothesis
    \[
    y_{t+1} \geq  y_t - \frac{1}{t+1}y_t^2
    \geq \frac{1}{t^{1/2}}y_1 - \frac{1}{t+1}\cdot\frac{y_1^2}{t} \geq \frac{1}{(t+1)^{1/2}}y_1.
    \]
    Therefore, we have shown that $y_t \geq \frac{1}{\sqrt{t}}y_1$ for every $t\in \nn$ as long as $y_1$ is sufficiently small, which could be assumed to be true since otherwise will yield our result directly. To conclude, we notice that $t\cdot y_t \geq t^{1/2}\cdot y_1 = t^{1/2}\cdot(1-x_1) $, which goes to infinity as $t\rightarrow \infty$. 
\end{proof}

\section{Proof of \Cref{thm:SA-regret-bound}}
\setcurrentname{Proof of \Cref{thm:SA-regret-bound}}
\label{app: proof-regret-bound}
We first restate the theorem here.
\saRegretBound*
In this section, we will emphasise that the meaning of step-index $t$ is the total number of bettors interacting with the market. Denote the probability that the bettor will bet on the "not happen" and "happen" sides as $\kappa_t := F(1-\priceL_t)$ and $\rho_t:= 1- F(\priceR_t)$ respectively. We consider the "happen" side, with probability $\rho_t$ the process is updated as
\[
\priceR_{t+1} = \priceR_t - \eta_{t+1}(h(\priceR_t) + M_{t+1}),
\]
where $h(\priceR_t) =  \priceR_t + G(\priceR_t)-\expect{\pt~|~\pt\geq \priceR_t}$ and $M_{t+1} =  \expect{\pt~|~\pt\geq \priceR_t} - \widehat{p_t}$. 
First, we need to identify the following fact. 
\begin{lemma}\label{app-lem: taylor}
    Under the assumptions of \Cref{thm:SA-convergence}, if $(\priceR^\sharp,\priceL^\sharp)$ is a local maximiser of $u(\priceR,\priceL)$, then, within a compact convex neighbourhood $\mathcal{K}$ of $\priceR^\sharp$, there exists constants $\alpha, \beta>0$ such that
    \[
    \alpha (\priceR - \priceR^\sharp)^2 \leq h(\priceR)\cdot(\priceR - \priceR^\sharp) \leq \beta(\priceR - \priceR^\sharp)^2.
    \] 
\end{lemma}
\begin{proof}
    By \citet[Lemma D.2]{mertikopoulos2020almost}, we first have
    \[
    -\beta' (\priceR - \priceR^\sharp)^2 \leq \frac{\partial u}{\partial \priceR}\cdot(\priceR - \priceR^\sharp) \leq -\alpha'(\priceR - \priceR^\sharp)^2,
    \]
    for some positive constants $\alpha', \beta'$. We also notice that
    \[
    \begin{aligned}
         \frac{\partial u}{\partial \priceR} &= \frac{(\prefCDF(\priceR) - 1)(\priceR -g)}{\priceR(1-\priceR)} + \frac{\expect{(\pt - \priceR)_+}(\priceR^2 - 2g\priceR +g)}{\priceR^2(1 - \priceR)^2} \\
         &= \frac{(1-F(\priceR))(\priceR^2 - 2g\priceR + g)}{\priceR^2(1-\priceR)^2}\left[-G(\priceR) + \expect{\pt~|~\pt\geq \priceR} - \priceR\right] \\
         &= \frac{(1-F(\priceR))(\priceR^2 - 2g\priceR + g)}{\priceR^2(1-\priceR)^2}\cdot (-h(\priceR)).
    \end{aligned}
    \]
    It is clear that the function $\frac{(1-F(\priceR))(\priceR^2 - 2g\priceR + g)}{\priceR^2(1-\priceR)^2}$ is strictly positive and continuous for $0<a<1$. Since the neighbourhood $\mathcal{K}$ is compact, hence the function is bounded. In particular, it is lower bounded by some positive constant as long as $\mathcal{K}$ is small enough. Therefore, the result follows.
\end{proof}

We are ready the present the descent lemma.

\begin{lemma}
Define $D_t = \frac{1}{2}(\priceR_t - \priceR^\sharp)^2$. Let $\mathcal{K}$ and $\alpha$ be as in \Cref{app-lem: taylor}. Suppose the process stays in $\mathcal{K}$ whenever $t \geq 1$. If the bettor's belief $p_{t+1} \geq \priceR_t$ then
\begin{equation}
D_{t+1} \leq (1-2\alpha\eta_{t+1})D_t + \eta_{t+1}\xi_{t+1} + \frac{1}{2}\eta_{t+1}^2\tilde{h}(\priceR_t)^2, \label{app-eq: descent-lemma}
\end{equation}
where $\xi_{t+1} =  M_{t+1}(\priceR^\sharp- \priceR_t)$ and $\tilde{h}(\priceR_t) = h(\priceR_t) + M_{t+1}$.\label{app-lem: descent-lemma}
\end{lemma}
\begin{proof}
    We could notice that,
    \[
    \begin{aligned}
        D_{t+1} &= \frac{1}{2}(\priceR_{t+1} - \priceR^\sharp)^2 \\
        &=\frac{1}{2}\left[\priceR_t - \eta_{t+1}(h(\priceR_t) + M_{t+1})- \priceR^\sharp\right]^2 \\
        &= D_t - \eta_{t+1}(h(\priceR_t) + M_{t+1})(\priceR_t - \priceR^\sharp) + \frac{1}{2}\eta_{t+1}^2(h(\priceR_t) + M_{t+1})^2 \\
        &\leq D_t -2\alpha\eta_{t+1}D_t + \eta_{t+1}\xi_{t+1} + \frac{1}{2}\eta_{t+1}^2(h(\priceR_t) + M_{t+1})^2,
    \end{aligned}
    \]
    where the last inequality follows from \Cref{app-lem: taylor}.
\end{proof}
Next, we present a technical lemma characterising the growth of the recurrence relation we are studying.
\begin{lemma}[{\citet[Lemma 2]{chung1954stochastic}}]
    Let $(x_t)_{t\in\nn}$ be a nonnegative sequence such that
    \[
    x_{t+1} \leq \left(1 - \frac{P}{t+m}\right)x_t + \frac{R}{(t+m)^{1+r}},
    \]
    where $r,m,P,R>0$ and $P>r$. Then
    \[
    x_t \leq \frac{R}{P-r}\frac{1}{t} + o\left(\frac{1}{t}\right).
    \] \label{app-lem: SA-regret-recurrence}
\end{lemma}
Now, we are ready to prove \Cref{thm:SA-regret-bound}.
\begin{proof}[Proof of \Cref{thm:SA-regret-bound}]
    We first assume that the process is bounded within the neighbourhood $\mathcal{K}$ of a maximiser $\priceR^\sharp$ defined in \Cref{app-lem: descent-lemma}. We note that
    \begin{equation}
    \expect{\eta_{t+1}\xi_{t+1}} = \eta_{t+1}\expect{\expect{\xi_{t+1}~|~\mathcal{F}_t}} = \eta_{t+1}\expect{(\priceR^\sharp- \priceR_t)\expect{M_{t+1}~|~\mathcal{F}_t}} = 0.
    \label{app-eq: SA-step-control-martingale-diff}
    \end{equation}
    And also,
    \begin{equation}
    \expect{\frac{1}{2}\eta_{t+1}^2\tilde{h}(\priceR_t)^2} = \frac{1}{2}\eta_{t+1}^2\expect{(h(\priceR_t) + M_{t+1})^2} \leq \eta_{t+1}^2\expect{h(\priceR_t)^2 +  M_{t+1}^2}\leq 5\eta_{t+1}^2,
    \label{app-eq: SA-step-control-squared-noise}
     \end{equation}
    where the last inequality follows from \Cref{app-lem: noise-bounded}. 
    Denote $A_t$ as the event that bettor $t$ will bet on the positive side. Then
    \[
    \begin{aligned}
        \expect{D_{t+1}} &= \rho_t\expect{D_{t+1}~|~A_{t+1}} + (1-\rho_t)\expect{D_{t+1}~|~A_{t+1}^c}\\
        &= \rho_t\expect{D_{t+1}~|~A_{t+1}} + (1-\rho_t)\expect{D_t}\\
        &\leq \rho_t(1 -2\alpha\eta_{t+1})\expect{D_t}  + 5\rho_t \eta_{t+1}^2 + (1-\rho_t)\expect{D_t} \\
        &= (1 -2\alpha\rho_t\eta_{t+1})\expect{D_t}+5\rho_t\eta_{t+1}^2,
    \end{aligned}
    \]
    where the third line follows from \Cref{app-lem: descent-lemma} and the fact that the sigma-algebras $\sigma(A_{t+1})$ and $\sigma(D_t)$ are independent. Further, since the process is bounded within $\mathcal{K}$, we have 
    \[
    \rho_t= 1 - F(\priceR_t) \geq 1 - F(\sup \mathcal{K}) =: \rho.
    \]
    Since $\mathcal{K}$ could be arbitrarily small, and the number of maximisers is finite, we could assume that $\sup \mathcal{K} < 1$ and hence $\rho > 0$. Then
    \[
    \expect{D_{t+1}} \leq (1 -2\alpha\rho\eta_{t+1})\expect{D_t}+5\eta_{t+1}^2
    \]
    Hence, by \Cref{app-lem: SA-regret-recurrence}, when $\alpha\rho\gamma >1$ and $t$ is large enough, we have
    \[
    \expect{D_t} \leq \frac{5}{2\alpha\rho\gamma-1}\frac{1}{t} + o\left(\frac{1}{t}\right) \leq \frac{6}{t},
    \]
    where $\gamma$ could be taken to be large enough that $\alpha\rho\gamma > 1$.  Recalling that $\eta_{t+1} = \frac{\gamma}{t + m}$, this does not affect the former statements on restricting the process within $[g,1]$ since we could let $m$ be large enough accordingly. By the Jensen's inequality,
    \[
    \expect{\left|\priceR_t - \priceR_t^\sharp\right|}^2 \leq 2\expect{D_t}\leq  \frac{12}{t},
    \]
    which implies that $\left|\expect{|\priceR_t - \priceR_t^\sharp|}\right| \leq \sqrt{\frac{12}{t}}$. 
    By the Lipschitz continuity of $u$, we could further get
    \[
    \expect{\left|u(\priceR_t,\priceL_t) - u(\priceR_t^\sharp,\priceL_t^\sharp)\right|} \leq L_u\expect{\left|\priceR_t - \priceR_t^\sharp\right| + \left|\priceL_t - \priceL_t^\sharp\right|}\leq 4\sqrt{3}L_ut^{-1/2}.
    \]
    Now we ease the assumption that the process $(\priceR_t)_{t=1}^\infty \subset \mathcal{K}$. By \Cref{thm:SA-convergence}, the process will converge to at least one of the maximisers with probability $1$. Hence, there exists some constant $N$, we must have $(\priceR_t)_{t=N}^\infty \subset \mathcal{K}$. Therefore, for $t$ large enough, we could eventually get 
    \[
    \expect{\left|u(\priceR_t,\priceL_t) - u(\priceR_t^\sharp,\priceL_t^\sharp)\right|} \leq 4\sqrt{3}L_u(t-N)^{-1/2} \leq 7L_ut^{-1/2}.
    \]
    Finally, as we could not make sure which maximiser is $(\priceR^\sharp,\priceL^\sharp)$, we could at least take $(\priceR^\sharp,\priceL^\sharp)$ as the worst maximiser as stated in the theorem, which yields the result.
\end{proof}

\section{Proof of \Cref{thm: SA-regret-conditioned}}
\setcurrentname{Proof of \Cref{thm: SA-regret-conditioned}}
\label{app: Proof-SA-regret-conditioned}
\saRegretConditioned*
Let $\mathcal{U}$ be a neighbourhood of $\priceR^\sharp$ and $\epsilon>0$, similar to \citet{mertikopoulos2020almost}, we will use the following notations,
\[
\begin{aligned}
    A_{t} &= \{p_{t} \geq p_{m,t-1}\}, \\
    \zeta_t &= \sum_{k=1}^t \mathds{1}_{A_{k+1}}\eta_{k+1}\xi_{k+1},~~ S_t = \frac{1}{2}\sum_{k=1}^t\mathds{1}_{A_{k+1}}\eta_{k+1}^2\tilde{h}(p_{m,k})^2,\\
    R_t &= \zeta_t^2 + S_t,\\
    \Omega_t &= \Omega_t(\mathcal{U}) = \{p_{m,k}\in \mathcal{U}~\text{for all}~k=1,2,\ldots,t\},\\
    E_t &= E_t(\epsilon) = \{R_k \leq \epsilon~\text{for all}~k=1,2,\ldots,t\}.\\
\end{aligned}
\]
In particular, we make the following relation between $\mathcal{U}$ and $\epsilon$ that
\[
\{\priceR:(\priceR - \priceR^\sharp)^2 \leq 4\epsilon + 2\sqrt{\epsilon}\} \subset \mathcal{U}.
\]
The following technical lemma further controls the noise terms $R_t$. We will present the proof for the sake of completeness but we note that the proof idea is similar to Lemma D.3 of \citet{mertikopoulos2020almost}.
\begin{lemma}[{\citet[Lemma D.3]{mertikopoulos2020almost}}]
    Asumme that $(\priceR_{1} - \priceR^\sharp)^2 \leq 2\epsilon$, under the assumptions in \Cref{thm:SA-convergence}, for $t = 1,2,\ldots$, we have
    \begin{enumerate}
        \item $\Omega_{t+1} \subset \Omega_t$ and $E_{t+1} \subset E_t$.
        \item $E_{t-1}\subset \Omega_t$.
        \item Consider the following event
        \[
        \tilde{E}_t := E_{t-1} \setminus E_t = E_{t-1}\cap \{R_t > \epsilon\} = \{R_k\leq\epsilon~\text{for all}~k=1,2,\ldots,n-1~\text{and}~R_t > \epsilon\},
        \]
        and let $\tilde{R}_t = R_t\mathds{1}_{E_{t-1}}$ denote the cumulative error subject to the fact that the noise is being small until $t$, then
        \begin{equation}
        \expect{\tilde{R}_t} \leq \expect{\tilde{R}_{t-1}} + \left(5 + r_\mathcal{U}^2\right)\eta_{t+1}^2 - \epsilon\prob{\tilde{E}_{t-1}},
        \label{app-eq: refined-no-escape}
         \end{equation}
        where $r_\mathcal{U} = \sup_{\priceR\in \mathcal{U}}|\priceR - \priceR^\sharp|$ and, by convention, we set $\tilde{E}_0 = \emptyset$ and $\tilde{R}_0 = 0$.
    \end{enumerate}
\end{lemma}
\begin{proof}
    For 1, it comes directly from the definition of $E_t$ and $\Omega_t$.
    
    For 2, we will prove it by induction. For the base case ($t=1$), we have $E_0 = \Omega$ which is the whole probability space, and $\Omega_1 = \{\priceR_{1} \in \mathcal{U}\}$ which is specified by the assumption that $(\priceR_{1} - \priceR^\sharp)^2 \leq 2\epsilon$. For the step case, we assume that $E_{t-1} \subset \Omega_t$. It is equivalent to the fact that if $R_k \leq \epsilon$ for every $k = 1,2,\ldots, t-1$ then $p_{m,k} \in \mathcal{U}$ for all $k = 1,2\ldots t$. 
    Hence, by the fact that $E_{t+1} \subset E_t$, we only need to show that $\priceR_{t+1} \in \mathcal{U}$ when $R_t \leq \epsilon$. 
    For realisations that $p_{t+1} < \priceR_t$, we could see that $\priceR_{t+1} = \priceR_t \in \mathcal{U}$. 
    For realisations that $p_{t+1} \geq \priceR_t$, by \Cref{app-lem: descent-lemma}, we have
    \[
    D_{k+1} \leq D_k + \eta_{k+1}\xi_{k+1} + \frac{1}{2}\eta_{k+1}^2\tilde{h}(\priceR_t)^2,
    \]
    for every $k = 1,2,\ldots,t$. Therefore, summing up both sides from $k=1$ to $k=t$ gives
    \[
    D_{t+1} \leq D_1 + \zeta_t + S_t \leq \sqrt{R_t} + R_t \leq \epsilon + \sqrt{\epsilon} + \epsilon = 2\epsilon + \sqrt{\epsilon}.
    \]
    Since $\{\priceR:(\priceR - \priceR^\sharp)^2 \leq 4\epsilon + 2\sqrt{\epsilon}\} \subset \mathcal{U}$, the results follows.

    For 3, we notice that
    \begin{equation}
    \begin{aligned}
    \tilde{R}_{t} &= R_t\mathds{1}_{E_{t-1}} =  R_{t-1}\mathds{1}_{E_{t-1}} + (R_t - R_{t-1})\mathds{1}_{E_{t-1}}\\
    &= R_{t-1}\mathds{1}_{E_{t-2}} - R_{t-1}\mathds{1}_{\tilde{E}_{t-1}} + (R_t - R_{t-1})\mathds{1}_{E_{t-1}} \\
    &= \tilde{R}_{t-1} + (R_t - R_{t-1})\mathds{1}_{E_{t-1}} - R_{t-1}\mathds{1}_{\tilde{E}_{t-1}}. \label{app-eq: refined-lemma-decomp}
     \end{aligned}
    \end{equation}
    We first focus on the term $(R_t - R_{t-1})\mathds{1}_{E_{t-1}}$. If $p_t < \priceR_t$, then $R_t - R_{t-1} = 0$. If $p_t \geq \priceR_t$, we have
    \[
    \begin{aligned}
    R_t - R_{t-1}&= \zeta_t^2 - \zeta_{t-1}^2 + S_t - S_{t-1} \\
    &= \eta_{t+1}\xi_{t+1}(\zeta_t + \zeta_{t-1}) + \frac{1}{2}\eta_{t+1}^2\tilde{h}(\priceR_t)^2 \\
    & = 2\eta_{t+1}\xi_{t+1}\zeta_{t-1} + \eta_{t+1}^2\xi_{t+1}^2 + \frac{1}{2}\eta_{t+1}^2\tilde{h}(\priceR_t)^2.
        \end{aligned}
    \]
    Then, we deal with the above term by term.
    For the first term, we have
    \[
    \expect{\mathds{1}_{E_{t-1}}2\eta_{t+1}\xi_{t+1}\zeta_{t-1}~|~A_t} = \expect{\mathds{1}_{E_{t-1}}2\eta_{t+1}\zeta_{t-1}\expect{\xi_{t+1}~|~\mathcal{F}_t}~|~A_t} = 0,
    \]
    where the first equality follows from the fact that $\zeta_{t-1}$ and $\mathds{1}_{E_{t-1}}$ are $\mathcal{F}_t$ measurable and $\eta_{t+1}$ is constant, the second equality follows from the assumption that $\expect{M_{t+1}~|~\mathcal{F}_t} = 0$.
    For the second term, we have
    \[
    \expect{\eta_{t+1}^2\mathds{1}_{E_{t-1}}\xi_{t+1}^2} \leq \eta_{t+1}^2\expect{\mathds{1}_{\Omega_t}M_{t+1}^2(\priceR^\sharp - \priceR_t)^2}  \leq \eta_{t+1}^2r^2_\mathcal{U}.
    \]
    For the third term, we have
    \[
    \frac{1}{2}\eta_{t+1}^2\expect{\tilde{h}(\priceR_t)^2}\leq \frac{1}{2}\eta_{t+1}^2\expect{2(M_{t+1}^2 + h(\priceR_t)^2)} \leq 5.
    \]
    Therefore, in summary, we have
    \[
    \expect{\mathds{1}_{E_{t-1}}(R_t - R_{t-1})} \leq \eta_{t+1}^2(5+ r^2_\mathcal{U}).
    \]
    Next, for the last term of \eqref{app-eq: refined-lemma-decomp}, we have
    \[
    \expect{R_{t-1}\mathds{1}_{\tilde{E}_{t-1}}} \geq \epsilon\expect{\mathds{1}_{\tilde{E}_{t-1}}} = \epsilon\prob{\tilde{E}_{t-1}},   
    \]
    where the first inequality follows from the definition of $\tilde{E}_{t-1}$. Putting everything together to \eqref{app-eq: refined-lemma-decomp} will yield the result.
\end{proof}

The following lemma controls the probability of escaping.

\begin{lemma}[{\citet[Proposition D.2]{mertikopoulos2020almost}}]
  Fix the tolerance level $\delta>0$, under the assumptions in \Cref{thm:SA-convergence} and \Cref{thm:SA-regret-bound}, we have
  \[
  \prob{E_t} \geq 1-\delta, ~~\forall t\in\nn.
  \] \label{app-lem: SA-non-escape-prob}
\end{lemma}
\begin{proof}
    First, we note that
    \[
    \prob{\tilde{E}_{t}} = \prob{E_{t-1}\cap \{R_t > \epsilon\}} = \expect{\mathds{1}_{E_{t-1}}\cdot \mathds{1}_{\{R_t > \epsilon\}}}\leq \expect{\mathds{1}_{E_{t-1}}\cdot\frac{R_t}{\epsilon}} = \expect{\tilde{R}_t}/\epsilon,
    \]
    where the inequality follows from the fact that $R_t/\epsilon > 1$ when $R_t > \epsilon$.
    Next, summing over both sides from $1$ to $t$ for \eqref{app-eq: refined-no-escape}, we get
    \[
    \expect{\tilde{R}_t} \leq \expect{\tilde{R}_0} + R_\star\sum_{k=1}^t \eta_{k+1}^2 - \epsilon\sum_{k=1}^t \prob{\tilde{E}_{k-1}},
    \]
    where $R_\star = 5 + r_\mathcal{U}^2$. Hence, we combining the above findings, we have
    \[
    \sum_{k=1}^t \prob{\tilde{E}_{k}} = \prob{\tilde{E}_{t}} + \sum_{k=1}^t \prob{\tilde{E}_{k-1}} \leq \frac{1}{\epsilon}\left(\expect{\tilde{R}_t} + \expect{\tilde{R}_0}- \expect{\tilde{R}_t} +  R_\star\sum_{k=1}^t \eta_{k+1}^2\right) =\frac{R_\star}{\epsilon}\sum_{k=1}^t \eta_{k+1}^2.
    \]
    Therefore, by controlling the learning rate small enough such that $\frac{R_\star}{\epsilon}\sum_{k=1}^t \eta_{k+1}^2 \leq \delta$, and the fact that $(\tilde{E}_{k})_{k=1}^t$ are disjoint, we could conclude that
    \[
    \prob{E_t} = \prob{\bigcap_{k=1}^t \tilde{E}_k^c} = 1 - \prob{\bigcup_{k=1}^t \tilde{E}_t} = 1 - \sum_{k=1}^t\prob{\tilde{E}_k} \geq 1 - \delta.
    \]
\end{proof}

Now, we are ready to put everything together.
\begin{proof}[Proof of \Cref{thm: SA-regret-conditioned}]
First, we have
\[
\prob{\Omega_\mathcal{U}} = \prob{\bigcap_{t=1}^\infty\Omega_t} = \inf_{t\in \nn}\prob{\Omega_t} \geq \inf_{t\in \nn}\prob{E_{t-1}}\geq 1- \delta,
\]
where the second-to-last inequality follows from the fact that $E_{t-1}\subset \Omega_t$ for every $t\in \nn$, and the last inequality follows from \Cref{app-lem: SA-non-escape-prob}.
Next, by \eqref{app-eq: SA-step-control-martingale-diff} and \eqref{app-eq: SA-step-control-squared-noise}, we have
\[
\expect{\mathds{1}_{\Omega_{n}}\left(\eta_{t+1}\xi_{t+1} + \frac{1}{2}\eta_{t+1}^2\tilde{h}(\priceR)^2\right)} \leq \expect{\eta_{t+1}\xi_{t+1} + \frac{1}{2}\eta_{t+1}^2\tilde{h}(\priceR)^2} \leq 5\eta_{t+1}^2.
\]
Hence, let $\bar{D}_{t} = \expect{\mathds{1}_{\Omega_t}D_t}$, $\rho = 1-\sup \mathcal{U}$, by \eqref{app-eq: descent-lemma}, we have
\[
\begin{aligned}
\bar{D}_{t+1} &=  \rho_t\expect{\bar{D}_{t+1}~|~A_{t+1}} + (1-\rho_t)\expect{\bar{D}_{t+1}~|~A_{t+1}^c}\\
&= \rho_t\expect{\bar{D}_{t+1}~|~A_{t+1}} + (1-\rho_t)\expect{\bar{D}_{t}~|~A_{t+1}^c}\\
&\leq \rho_t(1 -2\alpha\eta_{t+1})\bar{D}_{t}  + \expect{\mathds{1}_{\Omega_{n}}\left(\eta_{t+1}\xi_{t+1} + \frac{1}{2}\eta_{t+1}^2\tilde{h}(\priceR)^2\right)}+ (1-\rho_t)\bar{D}_{t} \\
&\leq (1 -2\rho\alpha\eta_{t+1})\bar{D}_{t}  + 5\eta_{t+1}^2.
\end{aligned}
\]
where the third line follows from \Cref{app-lem: descent-lemma}, the independence of $\sigma(A_{t+1})$ and $\sigma(D_t)$, and the fact that $\Omega_{t+1}\subset\Omega_t$. Therefore, by \Cref{app-lem: SA-regret-recurrence}, we have
\[
\bar{D}_t \leq \frac{5}{2\alpha\rho\gamma -1}\frac{1}{t} + o\left(\frac{1}{t}\right) \leq \frac{6}{t},
\]
whenever $\alpha\rho\gamma\geq 1$ and $t$ is large enough. Also, we have
\[
\expect{(\priceR_t - \priceR^\sharp)^2~|~\Omega_\mathcal{U}} \leq \frac{\expect{(\priceR_t - \priceR^\sharp)^2\mathds{1}_{\Omega_\mathcal{U}}}}{\prob{\Omega_\mathcal{U}}} \leq \frac{2}{1-\delta}\bar{D}_t \leq \frac{12}{1-\delta}\cdot\frac{1}{t}.
\]
By Jensen's inequality, we could conclude that
\[
\expect{\left|\priceR_t - \priceR^\sharp\right|~|~\Omega_\mathcal{U}} \leq \sqrt{\frac{12}{1- \delta}}\cdot t^{-1/2}.
\]
Hence, 
\[
\begin{aligned}
    \expect{\left|u(\priceR_t, \priceL) - u(\priceR^\sharp,\priceL^\sharp)\right|~|~\Omega_\mathcal{U}} &\leq 
L_u\expect{\left|\priceR_t - \priceR^\sharp\right|~|~\Omega_\mathcal{U}} + L_u\expect{\left|\priceL_t - \priceL^\sharp\right|~|~\Omega_\mathcal{U}}\\
&\leq 4\sqrt{\frac{3}{1- \delta}}L_u\cdot t^{-1/2} \leq 2\sqrt{6}L_u\cdot t^{-1/2}.
\end{aligned}
\]
\end{proof}

%% file: appendix/online-algorithm-fair.tex
\section{Proof of \Cref{thm:regret-bound-FTL}}
\setcurrentname{Proof of \Cref{thm:regret-bound-FTL}}
\label{app: regret-bound-online-algo-fair}
\regretBoundFTL*
First, it is straightforward to verify that $u(\priceR,1-\priceR)$ is concave \wrt $\priceR$. By equating the gradient to $0$, we could get the following closed-form expression of the unique maximiser
\begin{equation}
\priceR^\star = \frac{\sqrt{g\cdot \expect{\pt}}}{\sqrt{g\cdot \expect{\pt}} + \sqrt{(1-g)\cdot (1-\expect{\pt})}}. \label{app-eq: opt-price-fair}
\end{equation}

The remainder of the proofs comes from determining a martingale property for the accumulation of \( \widehat{\pt} \), utilising the Azuma-Hoeffding Inequality, and exploiting the local Lipschitz properties of functions. We first recall the Azuma-Hoeffding Inequality.

\begin{theorem}[{Azuma-Hoeffding Inequality}]
\label{app-lem:azuma-hoeffding}
Let \( (X_{t})_{t \in \mathbb{N} \cup \{ 0 \}} \) be a martingale \wrt filtration \( (\mathcal{F}_{t\in\mathbb{N} \cup \{ 0 \}}) \). Suppose that \( \vert X_{t} - X_{t-1} \vert \leq c_{t} \) for all \( t \in \mathbb{N} \) for non-negative \( (c_t)_{t \in \mathbb{N}}\). Then
\begin{equation*}
    \prob{\left \vert X_{T} - X_0 \right \vert > r} \leq 2 \exp \left( - \frac{2r^2}{\sum^{T}_{t=1} c_{t}^2} \right).
\end{equation*}
\end{theorem}

\begin{proof}[Proof of \cref{thm:regret-bound-FTL}]
Let \( \tau > 0 \). By unwinding \cref{eq:cumavg_belief_noclip}, we have
\[
\overline{\pt} = \frac{1}{t}\sum_{t'=1}^t \widehat{p_{t'}}.
\]
Let $X_0 = 0$, and $X_t = \sum_{t'=1}^t (\widehat{p_{t'}}- \expect{\pt})$. 
Note that $X_t = X_{t-1} + (\widehat{p_{t}}- \expect{\pt})$.
Since $\widehat{p_{t}}$ is an unbiased estimator of $\expect{\pt}$ under the filtration $\calF_{t-1}$,
the sequence $(X_t)_{t\in \nn\cup \{0\}}$ is a martingale w.r.t.~to its natural filtration $(\calF_t)_{t\in \nn\cup \{0\}}$.
Also, due to the assumption that $w_t$ is bounded almost surely, each $\widehat{p_{t'}}$ is bounded almost surely too.
Thus, for all sufficiently large $T$, we can apply the Azuma-Hoeffding inequality to derive that
\[
\prob{X_T > r} \le 2 \exp \left(-\Omega(r^2/T)\right).
\]
In particular, by setting $r = \Theta(\sqrt{T\log \frac{1}{\delta}})$, the RHS of the above inequality is at most $\delta$.
By noting that $\overline{p_T} = \frac{1}{T} X_T + \expect{\pt}$, we can conclude that
with probability $1-\delta$, $\overline{p_T}$ lies within the interval $\expect{\pt} \pm \Theta(T^{-1/2}\sqrt{\log \frac{1}{\delta}})$.
When $T$ is sufficiently large, this interval is a strict subset of the interval $[\tau,1-\tau]$, so the clipping in \cref{alg: FTL} is no longer effective.

Thus, $a_T$ updated by the algorithm using \cref{eq: opt-price-fair} lies within the interval $\psi(\expect{\pt}) \pm \calO(T^{-1/2}\sqrt{\log \frac{1}{\delta}})$,
since the function $\psi$ is Lipschitz continuous locally around $\expect{\pt}$.
Here, we use the assumption that both $g$ and $\expect{\pt}$ lie in the open interval $(\tau,1-\tau)$, to ensure that
$\priceR^\star = \psi(\expect{\pt})\in (\tau,1-\tau)$ too, and hence the interval $\priceR^\star\pm \calO(T^{-1/2}\sqrt{\log \frac{1}{\delta}})$ is a subset of $(\tau,1-\tau)$
for all sufficiently large $T$.

Finally, the theorem follows by using the local Lipschitz continuity around $\priceR^\star$ of the profit function.
\end{proof}